\documentclass[aps,prd,reprint,amsmath,amssymb,superscriptaddress,nofootinbib,eqsecnum]{revtex4-1}

\usepackage[T1]{fontenc}
\usepackage{newtxtext}
\usepackage[varvw]{newtxmath}

\usepackage{titlesec}
\titleformat{\section}{\bfseries\center\uppercase}{\thesection.~}{0.1em}{}
\titleformat{\subsection}{\bfseries\center}{\thesubsection.~}{0.1em}{}
\titleformat{\subsubsection}{\bfseries\itshape\center}{\thesubsubsection.~}{0.1em}{}
\titlespacing{\section}{0pt}{2em plus 0.1em minus 0.1em}{0.7em}
\titlespacing{\subsection}{0pt}{1.5em}{0.5em}
\titlespacing{\subsubsection}{0pt}{1.5em}{0.5em}

\let\startappendix\appendix
\renewcommand{\appendix}{%
\startappendix
\titleformat{\section}{\bfseries\center\uppercase}{APPENDIX~\thesection:~}{0em}{}
}

\makeatletter
\renewcommand\@makefntext[1]{%
\noindent{\hspace{10pt}}{\@makefnmark}#1}
\makeatother

\renewcommand{\footnoterule}{%
  \kern -3pt
  \hrule width 1.2cm
  \kern 4pt
}

\DeclareMathAlphabet{\mathcal}{OMS}{cmsy}{m}{n}

\usepackage[dvipsnames]{xcolor}
\usepackage[colorlinks=true,allcolors=Blue,breaklinks=true]{hyperref}

\usepackage{enumitem}
\setlist[enumerate]{topsep={0pt},label={(\arabic*)},after=\vspace{0pt},itemsep=0pt,parsep=0.2em}

\usepackage{graphicx}

\usepackage{bm,slashed}

\newcommand{\Mp}{\ensuremath{M_\text{Pl}}}
\newcommand{\ovl}{\overline}
\newcommand{\w}{\omega}
\newcommand{\bw}{\bar\w}
\newcommand{\avg}[1]{\left< #1 \right>} 

\begin{document}

\title{Constraining screened fifth forces with the electron magnetic moment}

\author{Philippe \surname{Brax}}
\email[]{philippe.brax@ipht.fr}
\affiliation{Institut de Physique Th\'eorique, Universit\'e Paris-Saclay, CEA, CNRS, F-91191 Gif/Yvette Cedex, France}

\author{Anne-Christine \surname{Davis}}
\email[]{A.C.Davis@damtp.cam.ac.uk}
\affiliation{DAMTP, Centre for Mathematical Sciences, University of Cambridge, Wilberforce Road, Cambridge CB3 0WA, United Kingdom}

\author{Benjamin \surname{Elder}}
\email[]{Benjamin.Elder@nottingham.ac.uk}
\affiliation{School of Physics and Astronomy, University of Nottingham, Nottingham NG7 2RD, United Kingdom}

\author{Leong Khim \surname{Wong}}
\email[]{L.K.Wong@damtp.cam.ac.uk}
\affiliation{DAMTP, Centre for Mathematical Sciences, University of Cambridge, Wilberforce Road, Cambridge CB3 0WA, United Kingdom}

\date{April 27, 2018}

\begin{abstract}
Chameleon and symmetron theories serve as archetypal models for how light scalar fields can couple to matter with gravitational strength or greater, yet evade the stringent constraints from classical tests of gravity on Earth and in the Solar System. They do so by employing screening mechanisms that dynamically alter the scalar's properties based on the local environment. Nevertheless, these do not hide the scalar completely, as screening leads to a distinct phenomenology that can be well constrained by looking for specific signatures. In this work, we investigate how a precision measurement of the electron magnetic moment places meaningful constraints on both chameleons and symmetrons. Two effects are identified: First, virtual chameleons and symmetrons run in loops to generate quantum corrections to the intrinsic value of the magnetic moment---a common process widely considered in the literature for many scenarios beyond the Standard Model. A second effect, however, is unique to scalar fields that exhibit screening. A scalar bubblelike profile forms inside the experimental vacuum chamber and exerts a fifth force on the electron, leading to a systematic shift in the experimental measurement. In quantifying this latter effect, we present a novel approach that combines analytic arguments and a small number of numerical simulations to solve for the bubblelike profile quickly for a large range of model parameters. Taken together, both effects yield interesting constraints in complementary regions of parameter space. While the constraints we obtain for the chameleon are largely uncompetitive with those in the existing literature, this still represents the tightest constraint achievable yet from an experiment not originally designed to search for fifth forces. We break more ground with the symmetron, for which our results exclude a large and previously unexplored region of parameter space. Central to this achievement are the quantum correction terms, which are able to constrain symmetrons with masses in the range $\mu \in [10^{-3.88},10^8]\,\text{eV}$, whereas other experiments have hitherto only been sensitive to 1 or 2 orders of magnitude at a time. 
\end{abstract}

\maketitle

\section{Introduction}
Many laboratory experiments exist today to search for or otherwise strongly constrain deviations from Newtonian gravity on submillimeter scales 
\cite{Adelberger:2003zx,Will:2014kxa,Burrage:2016bwy,Burrage:2017qrf}. These often give tight bounds on the parameters of hypothetical Yukawa fifth forces, although it has recently become interesting also to consider their implications for nonlinear scalar fields. It is now known that when a scalar field is allowed to have both self-interactions and nonlinear couplings to the Standard Model, its phenomenology becomes markedly different.

\subsection{Chameleonlike particles}

Despite the enormous range of possibilities (see \cite{Clifton:2011jh,Joyce:2014kja} for reviews), a  defining feature common to such scalar fields is a nonperturbative effect known as screening. Screening mechanisms drive the scalar to dynamically alter its properties in response to its surroundings, thus suppressing or enhancing the fifth force it mediates. Two models of screening are particularly suited to being tested in the laboratory and have justly been the focal point of experiments in recent years. The first is the chameleon mechanism \cite{Khoury:2003aq,Khoury:2003rn}, wherein the mass of the scalar varies accordingly with the ambient density, thus resulting in a Yukawa-like suppression of the range of its fifth force in dense environments. The second, dubbed the symmetron \cite{Hinterbichler:2010es,Hinterbichler:2011ca}, utilizes a Higgs-like potential and the spontaneous breaking of its $\mathbb Z_2$ symmetry to couple the scalar to matter when in high vacuum while decoupling it in dense media. Both models belong to the same universality class of scalar-tensor theories, and serve as archetypal examples of how variations in density can elicit screening. In this paper, we introduce the blanket term ``chameleonlike particle'' (CLP) to make it easier to refer to this class of models collectively.\footnote{Our choice of nomenclature draws inspiration from and highlights the contrast with axionlike particles (ALPs), which are (pseudo)scalar fields that do not couple to matter.}

At the time of its introduction, this novel idea of screening found tremendous success in enabling a CLP's evasion of the the stringent fifth force constraints enforced by tests of gravity on Earth and in the Solar System that were already in place \cite{Will:2014kxa}. However, in some sense this success has been its own demise; having spurred the onset of a number of dedicated experiments searching specifically for signatures of screening. Today, most of the parameter space of the original chameleon model has been ruled out, leaving only a sliver still out of reach of current experiments. (See Ref.~\cite{Burrage:2017qrf} for a review of current constraints on CLPs.) In contrast, the space of symmetron models remains mostly unexplored. This state of affairs is due primarily to a lack of theoretical work in translating bounds from existing experiments conducted for the chameleon, although some of the blame is also borne by the symmetron's distinct phenomenology. Many laboratory experiments conducted in vacuum chambers are only sensitive to a small range of the symmetron mass (discussed further in Sec.~\ref{sec:sym}), meaning a large number of complementary experiments are needed to probe the parameter space fully. All in all, the question of whether scalar fifth forces exist in our Universe still remains open today. Our aim in this paper is to make further progress in answering this question.

We do so by taking an approach complementary to dedicated searches: A small number of high-precision experiments conducted and refined over the years have verified the accuracy of the Standard Model, and QED in particular, to the level of about one part per trillion. As CLPs are assumed to interact with all matter species, if present, they can give rise to additional effects that might tarnish this spectacular agreement between experiment and theory. Theoretical work in reanalyzing precision QED tests while incorporating the effects of such scalar fields is therefore interesting, since models in conflict with known physics can immediately be deemed unviable. Moreover, such work is also useful in elucidating where in parameter space future searches should direct their focus.

\subsection{Anomalous magnetic moment}

In this work, we investigate how the precision measurement of the electron's magnetic moment places bounds on both chameleons and symmetrons. The magnetic moment $\bm\mu$ can be written as
\begin{equation*}
\bm\mu = - g \mu_B \mathbf{S}
\end{equation*}
in terms of the spin $\mathbf{S}$ and the Bohr magneton $\mu_B = e/2 m_e$. (We work in units with $\hbar = c = 1$ throughout.) In the current state of the art, what is measured experimentally is the dimensionless ratio $g/2$, which is exactly one for a classical field governed by the Dirac equation. As is well known, quantum fluctuations slightly increase this value, making it a promising probe for the existence of new physics. The difference between the true and tree-level values is called the anomalous magnetic moment
\begin{equation*}
a = (g-2)/2.
\end{equation*}

To measure this, Hanneke~\emph{et al.}~\cite{HannekePRL,HannekePRA} confine a single electron in a cylindrical Penning trap, within which an axial magnetic field and quadratic electrostatic potential are maintained. The value of $a$ can then be inferred by measuring the eigenfrequencies of the electron in this vacuum cavity. Three measurements are needed: The cyclotron frequency $\bw_c$, the anomaly frequency $\bw_a$, and the axial frequency $\bw_z$, from which one deduces \cite{HannekePRA}
\begin{equation}
\label{eq:a_exp}
a_\text{exp} = \frac{\bw_a - \bw_z^2/(2\bw_c)}{\bw_c + 3 \delta_\text{rel}/2 + \bw_z^2/(2\bw_c)} + \frac{\Delta g_\text{cav}}{2}.
\end{equation}
In this paper, we denote experimentally measured frequencies $\bw_i$ with an overline to distinguish them from their theoretical counterparts. These, along with other experimental details relevant to this work, are discussed further in Sec.~\ref{sec:cav}. Two other quantities are present in Eq.~\eqref{eq:a_exp}: A small shift $\delta_\text{rel}$ is necessary to include the leading relativistic correction, whereas $\Delta g_\text{cav}$ is put in by hand to account for systematics arising from the interaction between the electron and radiation modes in the cavity. These considerations yield a measurement of $g/2$ precise to 0.28 parts per trillion \cite{HannekePRL,HannekePRA}:
\begin{equation*}
(g/2)_\text{exp} = 1.001\,159\,652\,180\,73\,(28).
\end{equation*}

Just as spectacular an achievement is its agreement with the Standard Model, which predicts a theoretical value 
\begin{equation}
\label{eq:a_sm}
a_\text{SM} =  \sum_{n=1}^\infty C_n (\alpha/\pi)^n + a_\text{ew} + a_\text{had}.
\end{equation}
The first term is the asymptotic series arising from QED, calculations for which have now been completed up to $n=5$ loops \cite{Aoyama:2012wj,Aoyama:2014sxa,*PhysRevD.96.019901}. Also relevant at the experiment's level of precision are small contributions from the electroweak and hadronic sectors, encapsulated in the remaining two terms. (See Ref.~\cite{Giudice:2012ms} for a more in-depth discussion.) The series in Eq.~\eqref{eq:a_sm} takes as input a value for the fine-structure constant that must be determined experimentally. For this purpose, the most precise, independent determination of $\alpha$ comes from combining measurements of the Rydberg constant \cite{codata} and the ratio $h/m_\text{Rb}$ obtained from recoil experiments with rubidium atoms \cite{Clade:2006zz,Cadoret:2008st,Bouchendira:2010es}. These yield the value
\begin{equation*}
\alpha^{-1}(\text{Rb}) = 137.035\,999\,049\,(90),
\end{equation*}
with the uncertainty dominated by the measurement of $h/m_\text{Rb}$. Substituting this into Eq.~\eqref{eq:a_sm}, the end result is an agreement between theory and experiment at 1.7 standard deviations \cite{Aoyama:2014sxa},
\begin{equation}
\label{eq:a_compare}
a_\text{SM} - a_\text{exp} = (1.30 \pm 0.77)\times 10^{-12}.
\end{equation}
The $1\sigma$ uncertainty above is dominated by the errors accrued in measuring $h/m_\text{Rb}$.

\subsection{Effects from a CLP}
\label{sec:intro_effects}

If a CLP exists in our Universe, three additional effects come into play:
\begin{enumerate}
\item \emph{Quantum corrections}: Virtual chameleons and symmetrons run in loops, generating additional corrections to the QED vertex function. These slightly increase the intrinsic value of the electron's magnetic moment.

\item\emph{Cavity shift}: Nonlinear scalar fields invariably form a bubblelike profile inside vacuum cavities, thus exerting an additional fifth force on the electron. This shifts its eigenfrequencies by a small amount $\w_i \to \w_i + \delta\w_i$. Unlike the intrinsic change in~(1), this is a systematic effect coming from the experimental setup, which must be corrected for to obtain an accurate value of $a_\text{exp}$.

\item\emph{Charge rescaling}: Scalars that couple to the photon induce a field-dependent rescaling of the electron charge, or equivalently, of the fine-structure constant $\alpha \to \alpha(\phi)$ \cite{PhysRevD.25.1527,Sandvik:2001rv,Barrow:2011kr,Olive:2007aj,Brax:2009ey,Brax:2010uq,Brax:2013doa}. If the local values of $\phi$ present in the experiments used to determine $\alpha(\text{Rb})$ differ from that in the Penning trap, then $\alpha(\text{Rb})$ must be appropriately rescaled before being substituted into Eq.~\eqref{eq:a_sm}.
\end{enumerate}
All three effects add up to an overall deviation $\delta a$. Compatibility with Eq.~\eqref{eq:a_compare} requires that this must be constrained, at the $2\sigma$ level, to lie within
\begin{equation}
\label{eq:da_constraint}
| \delta a + 1.30 \times 10^{-12}| < 1.54 \times 10^{-12}.
\end{equation}

Contributions from both the quantum and cavity effects can be estimated by considering the experiment of Hanneke~\emph{et al.}~in isolation, but including the variation of the fine-structure constant requires, in addition, a good understanding of how the scalar behaves in the experimental setups leading to the value of $\alpha(\text{Rb})$. This is a far more involved task, which lies beyond the scope of this paper. For simplicity, we shall assume in what follows that the value of $\alpha$ is identical in all relevant experiments. This assumption is not expected to have a negative impact on our results. Considering only the first two effects is sufficient to provide conservative bounds on the model parameters, which can only be expected to improve once charge rescaling is properly taken into account. In fact, only the bound on the photon coupling has room for improvement; our constraints for the matter coupling are robust against charge rescaling since the relevant physics is independent of~$\alpha$.

\subsection{Outline of this paper}
The remainder of this paper is organized as follows: The details that go into quantifying the effect of quantum corrections and the cavity shift are discussed in Secs.~\ref{sec:quantum} and \ref{sec:cav}, respectively. Up to this point, the calculations are kept as general as possible, and will apply to any nonlinear scalar field with a canonical kinetic term, a self-interaction potential, and couplings to the Standard Model. The reader interested primarily in the punchline may prefer to jump directly to Sec.~\ref{sec:chm}. There, the calculations are completed by specializing to the chameleon model, and the constraints on parameter space are determined. The same process is repeated for the symmetron in Sec.~\ref{sec:sym}. We summarize in Sec.~\ref{sec:conclusions}.

\section{Quantum corrections}
\label{sec:quantum}

The scalar fields we consider couple universally to matter and mediate a fifth force. At the quantum level, virtual exchange of these scalars leads to additional loop corrections to the QED vertex function, in turn resulting in an increase in the intrinsic value of the electron's magnetic moment.

\subsection{Lagrangian}
\label{sec:L}

We begin this section by briefly reviewing the ingredients that constitute chameleon and symmetron models. Both belong to the same family of scalar-field theories governed by the Lagrangian\footnote{For the purposes of laboratory experiments, it suffices to work in flat space. See, e.g., the reviews in Refs.~\cite{Clifton:2011jh,Joyce:2014kja} for the covariant form of this action. Our metric signature is $(-,+,+,+)$.}
\begin{equation}
\label{eq:L}
\mathcal L = -\frac{1}{2}(\partial\phi)^2 - V(\phi) + \mathcal L_m(\Psi,\phi),
\end{equation}
where the Standard Model fields (denoted collectively by $\Psi$) and their couplings to $\phi$ are encapsulated in the third term $\mathcal L_m$. Massive fermions, such as the electron, obey the modified Dirac equation \cite{Brax:2010jk}
\begin{equation}
\label{eq:Dirac}
\mathcal L_m \supset \ovl\psi[i\slashed D - \Omega(\phi) m_e]\psi,
\end{equation}
where $D_\mu = \partial_\mu + i e A_\mu$ is the usual gauge-covariant derivative, but the mass term has picked up a dependence on the scalar via the conformal function\footnote{$\Omega(\phi)$ is often also called $A(\phi)$ elsewhere in the literature. In this paper, we reserve $A$ for referring to the electromagnetic gauge field.} $\Omega(\phi)>0$. To satisfy the weak equivalence principle, nonrelativistic fluids with a conserved density distribution $\rho$ couple to $\phi$ via a similar interaction
\begin{equation}
\mathcal L_m \supset -\Omega(\phi)\rho.
\end{equation}
A coupling to the electromagnetic sector is also possible, since one is not forbidden by symmetries \cite{Brax:2009ey,Brax:2010uq}. Here one has the freedom to specify a different coupling function $\varepsilon(\phi)>0$, which modifies the kinetic term of the photon to read
\begin{equation}
\label{eq:defVarepsilon}
\mathcal L_m \supset - \frac{1}{4}\varepsilon(\phi) F_{\mu\nu} F^{\mu\nu}.
\end{equation}

As both $\Omega(\phi)$ and $\varepsilon(\phi)$ introduce nonrenormalizable operators into the Lagrangian, these theories should be viewed as low-energy effective field theories (EFTs) valid only below some cutoff. Well within this regime, these models typically satisfy $\Omega(\phi) \approx 1$ and $\varepsilon(\phi) \approx 1$. For this reason, their phenomenology is more aptly framed in terms of the dimensionless coupling strengths
\begin{equation}
\beta_m(\phi) = \Mp\frac{\text{d}\log\Omega}{\text{d}\phi},
\quad
\beta_\gamma(\phi) = \Mp\frac{\text{d}\log\varepsilon}{\text{d}\phi},
\end{equation}
where $\Mp = (8\pi G_\text{N})^{-1/2}$ is the reduced Planck mass. These theories are most interesting when $\beta_m, \beta_\gamma \geq 1$, corresponding to interactions that are of gravitational strength or greater.

\subsection{Vertex corrections}
\label{sec:quantum_loops}

To compute loop corrections, let us consider quantum fluctuations $\chi = \phi - \avg\phi$ about the classical background field profile $\avg\phi$ in the cavity where $g/2$ is to be measured. As the electron remains very close to the center of the cavity (see Sec.~\ref{sec:cav_motion}), it suffices to take $\avg\phi \approx \phi_0$ to be a constant, where $\phi_0$ is the classical field value at the center.

We shall restrict ourselves to the one-loop level, which is sufficient for determining the leading effect. At this order, the only influence from $V(\phi)$ is a mass term for the $\chi$ field, with mass $m_0$ given by the second derivative
\begin{equation}
\label{eq:def_mphi}
m_0^2 = V_{\text{eff},\phi\phi}(\phi_0)
\end{equation}
evaluated at the center of the cavity.\footnote{CLPs suffer from the usual hierarchy problem, since heavy particles running in loops induce large corrections to the scalar's mass. Some fine tuning must be tolerated in these theories to keep the classical predictions reliable.} Linearizing Eqs.~\eqref{eq:Dirac} and \eqref{eq:defVarepsilon}, the interaction terms relevant at this order are \cite{Brax:2009ey,Brax:2009aw}

~\vspace{-1\baselineskip}
\begin{equation}
\label{eq:L_linear_couplings}
\mathcal L_m \supset - \left( \frac{\beta_m m_e}{\Mp} \right) \ovl\psi\psi \chi - \frac{1}{4}\left(\frac{\beta_\gamma}{\Mp}\right) \chi F_{\mu\nu} F^{\mu\nu},
\end{equation}
where we write $\beta_m \equiv \beta_m(\phi_0)$ and $\beta_\gamma \equiv \beta_\gamma(\phi_0)$ for brevity. Overall factors of $\Omega(\phi_0) \approx 1$ and $\varepsilon(\phi_0) \approx 1$ can be absorbed into a renormalization of the electron mass $m_e$ and charge $-e$, respectively.

\begin{figure}
\includegraphics[width=70mm]{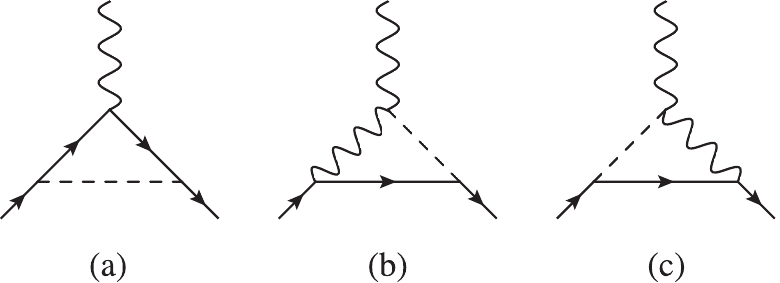}
\caption{Scalar field (dashed line) contributions at one-loop order to the magnetic moment of the electron.}
\label{fig:FeynmanDiagrams}
\end{figure}

Three Feynman diagrams contribute to the value of $g/2$ at one-loop order, as shown in Fig.~\ref{fig:FeynmanDiagrams}. As these diagrams have been widely considered for many different scenarios (see, e.g., Refs.~\cite{Giudice:2012ms,Jegerlehner:2009ry,PhysRevD.5.2396,Chen:2015vqy,Marciano:2016yhf}), we shall merely quote their result here in the main text. For the benefit of the inquisitive reader, a brief description of how these computations are carried out is relegated to Appendix~\ref{app:feyn}.

The first diagram in Fig.~\ref{fig:FeynmanDiagrams}(a) gives the finite contribution
\begin{equation}
\label{eq:da_bMbM}
\delta a \supset 2\beta_m^2 \left(\frac{m_e}{4\pi\Mp}\right)^2 I_1(m_0/m_e),
\end{equation}
whereas the remaining two diagrams are UV divergent. After renormalization in the $\overline{\text{MS}}$ scheme, they yield
\begin{equation}
\label{eq:da_bMbG}
\delta a \supset 4\beta_m\beta_\gamma\left(\frac{m_e}{4\pi\Mp}\right)^2
\left[ \log\left(\frac{\mu}{m_e}\right) + I_2(m_0/m_e)\right],
\end{equation}
where $\mu$ is an arbitrary energy scale. These results are expressed in terms of two integrals,
\begin{subequations}
\label{eq:feyn_integrals}
\begin{align}
\label{eq:feyn_integral_1}
I_1(\eta) &= \int_0^1\text{d}x \frac{(1-x)^2(1+x)}{(1-x)^2+x \eta^2},
\\
\label{eq:feyn_integral_2}
I_2(\eta) &= \int_0^1\text{d}x\int_0^1\text{d}y (x-1)\log[x^2 + (1-x)y \eta^2];
\end{align}
\end{subequations}
for which closed-form expressions can be found. For $\eta \geq 0$, we have
\begin{subequations}
\label{eq:feyn_integrals_closedform}
\begin{align}
I_1(\eta) =&\,
\frac{3}{2} - \eta^2 - \eta^2 (3 - \eta^2) \log \eta
\nonumber\\
&
-\eta (\eta^2 - 4)^{1/2} (\eta^2 - 1) \log \left( \frac{\eta}{2} + \sqrt{\frac{\eta^2}{4} - 1} \right),
\label{eq:feyn_integrals_closedform_1}
\\
I_2(\eta) =&\,
\frac{3}{2} - \frac{\eta^2}{6} + \frac{\eta^2}{6} (\eta^2 - 6)\log\eta
\nonumber\\
&+ 
\frac{\eta}{6} (\eta^2 - 4)^{3/2}\log\left( \frac{\eta}{2} - \sqrt{\frac{\eta^2}{4} - 1} \right),
\label{eq:feyn_integrals_closedform_2}
\end{align}
\end{subequations}
where the principal branch should be taken when $\eta < 2$. Alternatively, a piecewise expression for $I_1$ can also be found in Ref.~\cite{Chen:2015vqy}. Most of the time, however, we shall find ourselves working in the regime $m_0 \ll m_e$, such that it suffices to set $m_0/m_e = 0$ in the integrals. Both then evaluate to
\begin{equation*}
I_1(0) = I_2(0) = \frac{3}{2}.
\end{equation*}

\subsection{Nonrenormalizability}
\label{sec:quantum_ren}

It is worth discussing the result in Eq.~\eqref{eq:da_bMbG} in more detail. The scalar-photon coupling $\chi F_{\mu\nu}F^{\mu\nu}$ is a dimension-five operator, whose inclusion renders the theory nonrenormalizable. This plagues the evaluation of the diagrams in Figs.~\ref{fig:FeynmanDiagrams}(b) and \ref{fig:FeynmanDiagrams}(c), as their UV-divergent parts cannot be renormalized into any of the existing parameters in the Lagrangian we started with, such as the electron charge or particle masses. This is not uncommon in a low-energy EFT, and it must be understood that the scalar-photon coupling cannot remain pointlike up to arbitrarily high energies. This is dealt with in Ref.~\cite{Marciano:2016yhf} by assuming a sharp momentum cutoff. Here, we shall take an alternative route compatible with dimensional regularization, although in practice the end results are similar, since physics should not depend on the choice of regulator.

The resolution is to recognize that under RG flow, the heavy degrees of freedom that have integrated out to generate the scalar-photon coupling must also generate a bare term\footnote{We have written the coupling as $a_0 \mu_B$ to make manifest its contribution to the magnetic moment. Of course, in an EFT language, one should think of this as $a_0\mu_B \sim c_5/M_\star$, where $c_5$ is a dimensionless coupling and $M_\star$ is the appropriate cutoff scale.}
\begin{equation}
\mathcal{L} \supset - a_0 \mu_B \ovl\psi S^{\mu\nu} F_{\mu\nu} \psi
\end{equation}
in the Lagrangian, where $S^{\mu\nu} = \frac{i}{4}[\gamma^\mu,\gamma^\nu]$. The UV divergences that arise at one loop can now be absorbed into counterterms that renormalize $a_0$. This   naturally gives an extra contribution $\delta a \supset a_0$, which, in the absence of knowledge of the UV completion, is a new parameter to be constrained by experiment. For simplicity, we shall assume the UV completion is such that $a_0$ is much smaller than the one-loop contributions in Eqs.~\eqref{eq:da_bMbM} and \eqref{eq:da_bMbG} that it can be safely neglected.

On the other hand, the arbitrary scale $\mu$ should in principle be fixed by measuring $g/2$ at a given energy, after which Eq.~\eqref{eq:da_bMbG} dictates how this changes as we vary the energy of the experiment. Unlike particle colliders, however, there is an ambiguity in determining the scale $\mu$ of low-energy experiments like the one considered in this paper. Nevertheless, as $\mu$ appears only as the argument of a logarithm, its exact value is not crucial, and in practice a conservative estimate is to set
\begin{equation*}
\log(\mu/m_e) \sim 1.
\end{equation*}

\section{Cavity shift}
\label{sec:cav}

A defining feature of CLPs is their predisposition for forming a bubblelike profile when trapped in a vacuum cavity. This nontrivial profile will couple to the electron confined to the center of the Penning trap, exerting a fifth force which mildly shifts the energies of the electron's eigenstates. Unlike the intrinsic change described in Sec.~\ref{sec:quantum}, this is a systematic effect arising from considerations of how the experiment is conducted, which can also be used to place constraints. In this section, we describe how to account for this cavity shift, and quantify its contribution to the total deviation $\delta a$. Details of the experiment are described along the way, when needed, but only at a cursory level sufficient for our analysis. We refer the interested reader to the original experimental papers \cite{HannekePRL,HannekePRA} or the associated review \cite{Brown} for a more comprehensive account.

\subsection{Vacuum cavity profile}
\label{sec:cav_profile}

The electron's magnetic moment is measured using what is called a one-electron quantum cyclotron. In this setup, a single electron is trapped in a cylindrical vacuum cavity of radius $r_0$ and half-height $z_0$. The values of all experimental parameters, and the measured frequencies, are curated in Table~\ref{table:values}. A uniform magnetic field
\begin{subequations}
\label{eq:bare_em_fields}
\begin{equation}
\label{eq:B}
\mathbf{B} = B_0 \hat{\mathbf{z}}
\end{equation}
is established within the cavity to split the energy levels of the electron's spin states. A quadratic electrostatic potential\footnote{This expression differs by an overall sign from Ref.~\cite{Brown} because we use the convention that the electron has charge $-e$. In our case, both constants $e$ and $V_0$ are positive.}
\begin{equation}
\label{eq:V}
V = \frac{V_0}{2d^2}\left( \frac{r^2}{2} - z^2 \right)
\end{equation}
\end{subequations}
is also present to keep the electron close to the center, where the constant
\begin{equation*}
d = (r_0^2/4 + z_0^2/2)^{1/2} \approx 3.5\,\text{mm}
\end{equation*}
can be thought of as a characteristic length scale of the trap.

\begin{table}
\caption{Values of the experimental parameters and frequencies, reproduced from Refs.~\cite{HannekePRL,HannekePRA}. Up to small differences, the theoretical frequencies $\{ \w_+, \w_0, \w_z \}$ are approximately related to their experimentally measured counterparts by $\w_+ \approx \w_0 \approx \bw_c$ and $\w_z \approx \bw_z$. (See text in Secs.~\ref{sec:cav_H0} and \ref{sec:cav_dH} for details.)}
\label{table:values}
{\renewcommand{\arraystretch}{0}
\begin{ruledtabular}
\begin{tabular}{lccc}
Magnetic field & $B_0$ & 5.36 & T \\
Electrode potential difference & $V_0$ & 101.4 & V \\
Cavity radius & $r_0$ & 4.5 & mm\\
Cavity height & $2z_0$ & 7.7 & mm\\[5pt]
Cyclotron frequency    & $\bw_c/2\pi$ & 150    & GHz \\ 
Anomaly frequency      & $\bw_a/2\pi$ & 174    & MHz \\
Axial frequency        & $\bw_z/2\pi$ & 200    & MHz \\
Magnetron frequency    & $\w_-/2\pi$  & 133    & kHz \\[5pt]
\end{tabular}
\end{ruledtabular}}
\end{table}

The profile of the scalar inside the vacuum cavity is determined by solving its field equation in the static limit,
\begin{equation}
\label{eq:eom_scalar}
\nabla^2\phi = V_{\text{eff},\phi},
\end{equation}
where the comma on the rhs denotes a derivative. It follows from the Lagrangian in Sec.~\ref{sec:L} that the effective potential differentiates to give
\begin{equation}
V_{\text{eff},\phi} = V_{,\phi} + \frac{\beta_m(\phi)\rho}{\Mp} + \frac{\beta_\gamma(\phi)\rho_\text{em}}{\Mp}.
\end{equation}

The electromagnetic energy density $\rho_\text{em} = (\mathbf{B}^2 - \mathbf{E}^2)/2$ that enters on the rhs is given by Eq.~\eqref{eq:bare_em_fields} in the interior of the cavity, while it can be assumed that it is unappreciable in the exterior. The distribution $\rho$ of matter is assumed to be piecewise constant, such that
\begin{equation*}
\rho =
\begin{cases}
\rho_\text{cav} & \text{inside the cavity } (r < r_0, |z| < z_0), \\
\rho_\text{wall} & \text{in the surrounding walls}. 
\end{cases}
\end{equation*}
While no direct measurement of the density of gas $\rho_\text{cav}$ in the cavity has been made, an estimate from a similar trap design places an upper bound on the number density of atoms at $100\,\text{cm}^{-3}$  \cite{HannekePRA,PhysRevLett.65.1317}. Assuming this remains true for the current implementation, and taking the average mass of a molecule to be that of nitrogen, we estimate
\begin{equation*}
\rho_\text{cav} \lesssim 5 \times 10^{-18}\,\text{kg}\,\text{m}^{-3}.
\end{equation*}
On the other hand, the trap electrodes and vacuum container surrounding the cavity are composed primarily of silver, quartz, titanium, and molybdenum \cite{HannekePRA}, which have typical densities
\begin{equation*}
\rho_\text{wall} \gtrsim 3 \times 10^3\,\text{kg}\,\text{m}^{-3}.
\end{equation*}

For the two-dimensional cylindrical geometry considered here, an analytic solution to Eq.~\eqref{eq:eom_scalar} is not known. We postpone a full numerical solution of this equation to Secs.~\ref{sec:chm} and \ref{sec:sym}, where we specialize to chameleon and symmetron models, respectively. Nevertheless, we can continue to make analytic progress in this section because the experiment is cooled to an extremely low temperature $T \sim 100\,\text{mK}$, such that the electron remains very close to the center of the cavity. (We shall be more quantitative about this in Sec.~\ref{sec:cav_motion}.) Whatever the field profile is, it can be Taylor expanded about the center, which we take to be the origin, as
\begin{equation}
\label{eq:phi_Taylor}
\phi \simeq \phi_0 + \phi_{rr} \frac{r^2}{2 r_0^2} + \phi_{zz} \frac{z^2}{2z_0^2}.
\end{equation}
The central field value $\phi_0$ is a local maximum, hence we must have $\phi_{rr},\phi_{zz}<0$. Reflection symmetry in all three spatial directions ensures that the expansion contains only even powers of $r$ and $z$. Quartic and higher-order terms have been neglected since they are suppressed by additional powers of $\langle r^2/r_0^2 \rangle \ll 1$ and $\langle z^2/z_0^2 \rangle\ll 1$.

\subsection{Electromagnetic corrections}
\label{sec:cav_em}

The coupling function $\varepsilon(\phi)$ should be thought of as a relative permittivity of the vacuum, since it appears in the Maxwell equations as
\begin{equation}
\partial_\nu (\varepsilon F^{\mu\nu}) = J^\mu.
\end{equation}
The presence of a nontrivial scalar profile $\phi$ polarizes the vacuum, generating bound charges and currents that go on to source corrections to the bare electromagnetic fields. In a previous paper \cite{Wong:2017jer}, two of us showed that, at least in the case of the spectral lines of hydrogenlike atoms, this effect is large enough that it must be included. Moreover, it led to terms that allow a constraint on $\beta_\gamma$ independently of $\beta_m$. Given the large magnetic field in the cavity, it is worth exploring if the same is true for this experiment.

Solving Maxwell's equations perturbatively in the Lorenz gauge, the first-order corrections are given by
\begin{equation}
\nabla^2 \delta A_\mu = \frac{\beta_\gamma(\phi)}{\Mp} F_{\mu\nu}^{(0)} \partial^\nu\phi,
\end{equation}
where $F_{\mu\nu}^{(0)}$ describes the bare (zeroth-order) electric and magnetic fields, as given in Eq.~\eqref{eq:bare_em_fields}. Restricting ourselves to the quadratic terms in Eq.~\eqref{eq:phi_Taylor}, the correction to the electrostatic potential is
\begin{equation}
\delta V = \frac{V_0}{2d^2}\frac{\beta_\gamma(\phi_0)}{\Mp}\left( \phi_{rr} \frac{r^4}{16 r_0^2} - \phi_{zz} \frac{z^4}{6 z_0^2}\right),
\end{equation}
whereas the magnetic field receives corrections of the form
\begin{subequations}
\begin{align}
\delta\mathbf{A} &= B_0\phi_{rr}\frac{\beta_\gamma(\phi_0)}{\Mp}\frac{r^2}{8 r_0^2} (y\hat{\mathbf{x}} - x \hat{\mathbf{y}}),
\\
\delta\mathbf{B} &= - B_0\phi_{rr}\frac{\beta_\gamma(\phi_0)}{\Mp}\frac{r^2}{2 r_0^2} \hat{\mathbf{z}}.
\end{align}
\end{subequations}

\subsection{Hamiltonian}
\label{sec:cav_H}

The electron at the center of the Penning trap is adequately described by nonrelativistic quantum mechanics. In this limit, the modified Dirac equation in Eq.~\eqref{eq:Dirac} reduces to the Schr\"odinger equation with Hamiltonian \cite{Brax:2010jk,Brax:2010gp}
\begin{equation}
\label{eq:H}
H = \frac{(\mathbf{p} + e\mathbf{A})^2}{2m_e} - eV + g \mu_B \mathbf{B}\cdot\mathbf{S} + \Omega(\phi) m_e,
\end{equation}
where subleading terms of the form $\sim\mathcal O(\Omega\mathbf{p}^2)$ have been discarded. Ignoring the constant mass term, this Hamiltonian can be split into two parts,
\begin{equation*}
H = H_0 + \delta H.
\end{equation*}
The unperturbed Hamiltonian, for which the eigenstates can be determined exactly, is
\begin{equation}
\label{eq:H0_raw}
H_0 = \frac{\bm\pi^2}{2m_e} - eV + g \mu_B \mathbf{B}\cdot\mathbf{S},
\end{equation}
where the mechanical momentum is defined as $\bm\pi = \mathbf{p} + e \mathbf{A}$. It should be understood that the electromagnetic fields appearing here take their bare values, as in Eq.~\eqref{eq:bare_em_fields}. We work in the gauge $\mathbf{A} = (\mathbf{B} \times \mathbf{x})/2$. The remaining terms, which we shall treat with linear perturbation theory, are
\begin{equation}
\label{eq:cav_dH}
\delta H = \frac{m_e}{\Mp}\beta_m\delta\phi - e \delta V + \mu_B(2 \bm\pi \cdot\delta\mathbf{A} + g \delta\mathbf{B}\cdot\mathbf{S}).
\end{equation}
We write $\delta\phi$ to mean the quadratic terms in Eq.~\eqref{eq:phi_Taylor}, have resumed writing $\beta_m \equiv \beta_m(\phi_0)$ and $\beta_\gamma \equiv \beta_\gamma(\phi_0)$ for brevity, and have once again absorbed factors of $\Omega(\phi_0)$ into the electron mass $m_e$ (see Sec.~\ref{sec:quantum_loops}).

\subsection{Unperturbed eigenstates}
\label{sec:cav_H0}

The unperturbed Hamiltonian in Eq.~\eqref{eq:H0_raw} can be split into three mutually-commuting parts,
\begin{equation*}
H_0 = H_r + H_z + H_s.
\end{equation*}
The radial, axial, and spin interaction parts are, respectively,
\begin{subequations}
\begin{align}
H_r &= \frac{1}{2m_e}(\pi_x^2 + \pi_y^2) - \frac{1}{4} m_e \w_z^2 r^2,
\\
H_z &= \frac{1}{2m_e}\pi_z^2 + \frac{1}{2} m_e \w_z^2 z^2,
\\
H_s &= \frac{g}{2} \w_0 S_z.
\end{align}
\end{subequations}
These expressions are written in terms of the (bare) cyclotron frequency $\w_0$ and the axial frequency  $\w_z$, given by
\begin{equation}
\w_0 = e B_0 /m_e, \quad
\w_z = (e V_0/ m_e d^2)^{1/2}.
\end{equation}

It should already be clear at this stage that the axial motion, governed by $H_z$, simply corresponds to a harmonic oscillator with frequency $\w_z$. Making the transformation
\begin{equation}
z = \frac{1}{\sqrt{2m_e \w_z}}(a_z + a_z^\dagger), \quad
\pi_z = -i \sqrt{\frac{m_e \w_z}{2}}(a_z - a_z^\dagger)
\end{equation}
allows us to write
\begin{equation}
H_z = \w_z \left(a_z^\dagger a_z + \frac{1}{2}\right)
\end{equation}
in terms of creation and annihilation operators. It turns out that the same is true for the radial motion, which can be diagonalized to form two decoupled oscillators. To see this, first define two more frequencies $\w_\pm$ via \cite{Brown}
\begin{equation}
\label{eq:def_wpm}
2 \w_\pm = \w_0 \pm (\w_0^2 - 2\w_z^2)^{1/2},
\end{equation}
and denote their difference by $\Delta\w = \w_+ - \w_-$. Then, by writing
\begin{align}
x &= \frac{i}{\sqrt{2 m_e \Delta\w}}(a_c - a_c^\dagger + a_m - a_m^\dagger),
\nonumber\\
y &= - \frac{1}{\sqrt{2 m_e \Delta\w}}(a_c + a_c^\dagger - a_m - a_m^\dagger),
\nonumber\\
\pi_x &= \sqrt{\frac{m_e}{2 \Delta\w}}[\w_+(a_c + a_c^\dagger) - \w_-(a_m + a_m^\dagger)],
\nonumber\\
\pi_y &= i \sqrt{\frac{m_e}{2\Delta\w}}[\w_+(a_c - a_c^\dagger) + \w_-(a_m - a_m^\dagger)],
\end{align}
we ultimately end up with
\begin{align}
\label{eq:H0}
H_0 =&\; \w_+ \left( a_c^\dagger a_c + \frac{1}{2} \right) + \w_z \left(a_z^\dagger a_z + \frac{1}{2}\right) - \w_-\left(a_m^\dagger a_m + \frac{1}{2}\right)
\nonumber\\
& + \frac{g}{2} \w_0 S_z.
\end{align}
An eigenstate of this system $| n_c,n_z,n_m,m_s \rangle$ is specified by four quantum numbers: Three of these correspond to the occupation numbers $n_i = \langle a_i^\dagger a_i \rangle = 0,1,2,\dots$ of the harmonic oscillators, whereas the fourth is the spin state $m_s = \pm 1/2$.

Physically, the oscillators with frequencies $\{ \w_+, \w_z, \w_- \}$ correspond to cyclotron, axial, and magnetron motion, respectively (see Sec.~II of Ref.~\cite{Brown} for further details). That $\w_+$ is slightly larger than the bare cyclotron frequency $\w_0$ is due to the confining effect of the electrostatic potential, and note the minus sign appearing in front of $\w_-$ in Eq.~\eqref{eq:H0} makes clear that magnetron motion is unstable and unbounded from below. Based on the parameters of the experiment (see Table~\ref{table:values}), these frequencies satisfy the hierarchy
\begin{equation}
\label{eq:hierarchy}
\w_+ \gg \w_z \gg \w_-.
\end{equation}
 
\subsection{Axial and magnetron motion}
\label{sec:cav_motion}

This large hierarchy ensures that both the axial and magnetron motions are semiclassical. When measurements of the anomalous and cyclotron frequencies are being made, the axial motion is in thermal equilibrium with the detection amplifier circuit at a temperature $T_z \sim 230\,\text{mK}$ \cite{HannekePRA}. The average axial quantum number is thus given by
\begin{equation*}
n_z \sim k_B T_z/\w_z \sim 24.
\end{equation*}

Similarly, the magnetron motion thermalizes with a temperature $T_m \sim - (\w_-/\w_z) T_z$, assuming maximum axial sideband cooling \cite{HannekePRA,Brown}. This relation sets the axial and magnetron quantum numbers equal to each other,
\begin{equation*}
n_m \sim n_z \sim 24.
\end{equation*}
The negative temperature here again represents the fact that magnetron motion is unstable. Nevertheless, its decay time is on the order of billions of years, such that the state is metastable on the timescale of the experiment \cite{HannekePRA,Brown}.

These estimates justify us truncating the scalar field profile to quadratic order in Eq.~\eqref{eq:phi_Taylor}. For $n_c \sim 1$, the expectation values

\noindent
\begin{minipage}{\columnwidth}
{~}
{\vskip -1.7\baselineskip}
\begin{subequations}
\begin{align}
\avg{\frac{r^2}{r_0^2}} &= \frac{2(n_c + n_m + 1)}{m_e \Delta\w r_0^2} \sim 10^{-10},
\\
\avg{\frac{z^2}{z_0^2}} &= \frac{n_z + 1/2}{m_e \w_z z_0^2} \sim 10^{-7}
\end{align}
\end{subequations}
\medskip
\end{minipage}
\vskip -0.5em
\noindent
demonstrate that the spread of the electron wavefunction indeed remains very close to the center of the cavity.

\subsection{Frequency shifts}
\label{sec:cav_dH}

Three frequencies must be measured experimentally to determine the electron's magnetic moment. These are defined as follows:
\begin{subequations}
\label{eq:def_w}
\begin{enumerate}
\item The measured cyclotron frequency $\bw_c$ is obtained by exciting the electron from the state $(n_c,m_s) = (0,1/2) \to (1,1/2)$ at fixed $n_z$ and $n_m$. Taking the difference in the expectation values $\avg{H}$ for these two states, we get
\begin{equation}
\label{eq:def_w_c}
\bw_c = \w_+ - \frac{3}{2}\delta_\text{rel} + \delta\w_c.
\end{equation}
Note that the scalar-induced shift $\delta\w_c$ refers to the terms arising from computing $\avg{\delta H}$ at first order. Explicit expressions for all $\delta\w_i$ are given together below in Eq.~\eqref{eq:def_dw}. 
In Eq.~\eqref{eq:def_w_c}, we have also added in by hand the leading relativistic correction $\delta_\text{rel}/\w_+ \approx 10^{-9}$ relevant at the experimental precision \cite{HannekePRA,Brown}.

\item The measured anomaly frequency $\bw_a$ is similarly obtained by the excitation $(n_c,m_s) = (1,-1/2) \to (0,1/2)$. This yields
\begin{equation}
\bw_a = \frac{g}{2} \w_0 - \w_+ + \delta\w_a.
\end{equation}

\item The measured axial frequency $\bw_z$ corresponds to the transition $|\Delta n_z| = 1$, with all other quantum numbers fixed. This yields
\begin{equation}
\bw_z = \w_z + \delta\w_z.
\end{equation}
While the result does not change significantly, for definiteness we define $\bw_z$ as being the average energy for the two transitions $n_z \to n_z \pm 1$.
\end{enumerate}
\end{subequations}

\begin{widetext}
The three scalar-induced shifts are
\begin{subequations}
\label{eq:def_dw}
\begin{align}
\delta\w_c &= \frac{\phi_{rr}}{\Mp r_0^2}
\left[
\frac{\beta_m}{\Delta\w} - \frac{\beta_\gamma \w_0}{2 m_e \Delta\w}\left(\frac{g}{2} + (2n_m + 3)\frac{\w_+}{\Delta\w}\right)
+ (n_m + 1)\frac{\beta_\gamma \w_z^2}{2 m_e \Delta\w^2} - (2n_m+1)\frac{\beta_\gamma\w_0 \w_-}{2 m_e \Delta\w^2}
\right],
\\
\label{eq:def_dw_a}
\delta\w_a &= -\frac{\phi_{rr}}{\Mp r_0^2}
\left[
\frac{\beta_m}{\Delta\w} + (2n_m+3) \frac{\beta_\gamma \w_0}{2 m_e \Delta\w}\left(\frac{g}{2}-\frac{\w_+}{\Delta\w}\right)
+ (n_m + 1)\frac{\beta_\gamma \w_z^2}{2 m_e \Delta\w^2} - (2n_m+1)\frac{\beta_\gamma\w_0 \w_-}{2 m_e \Delta\w^2}
\right],
\\
\delta\w_z &= \frac{\phi_{zz}}{\Mp z_0^2}
\left[
\frac{\beta_m}{2\w_z} - (2n_z + 1)\frac{\beta_\gamma}{8m_e}
\right].
\end{align}
\end{subequations}
\newpage
\end{widetext}

At the moment, Eqs.~\eqref{eq:def_w} and \eqref{eq:def_dw} form a set of three simultaneous equations that relate $g/2$ to the measured frequencies $\bw_i = (\bw_c, \bw_a, \bw_z )$ and the theoretical parameters $\w_i = (\w_0, \w_z, \w_+,\w_- )$. We infer the value of the magnetic moment by eliminating all instances of $\w_i$ to obtain an expression for $g/2$ that depends only on $\bw_i$. This is necessary since $\bw_i$ are the only quantities measured to a high enough precision. To do so requires two more independent equations. These are provided by the definitions of $\w_\pm$ in Eq.~\eqref{eq:def_wpm}, which can be rearranged to read
\begin{equation}
\label{eq:w_relations}
\w_0 = \w_+ + \w_-, \quad
\w_- = \w_z^2/(2\w_0).
\end{equation} 
Note that these relations are exact for an ideal Penning trap, but are also approximately true in the presence of small imperfections of a real trap due to the hierarchy of Eq.~\eqref{eq:hierarchy} and an invariance theorem \cite{PhysRevA.25.2423}.

This set of five simultaneous equations will yield an approximate solution of the form
\begin{equation*}
(g/2)_\text{exp} = 1 + a_\text{exp} + \delta a_\text{cav},
\end{equation*}
where the zeroth-order term $a_\text{exp}$ is independent of the CLP, while the scalar-induced effects are encapsulated in the first-order correction $\delta a_\text{cav}$.  Owing to the highly nonlinear dependence of Eq.~\eqref{eq:def_dw} on $\w_i$, the desired result is most easily obtained in two stages. First, we solve this set of simultaneous equations at zeroth order by ignoring the scalar-induced shifts $\delta\w_i$. This is easy enough and returns $(g/2)_\text{exp} = 1 + a_\text{exp}$, with $a_\text{exp}$ given unsurprisingly by Eq.~\eqref{eq:a_exp} as before.

We then reintroduce the frequency shifts $\delta \w_i$ by perturbing $a_\text{exp}$ to first order to obtain the `cavity shift'\footnote{The minus sign is crucial, and reflects the fact that $\w_i$ are still the parameters to be eliminated. It can most easily be traced back to seeing that Eqs.~\eqref{eq:def_w} can be rearranged such that their lhs's read $\bw_i - \delta\w_i$.}
\begin{equation}
\delta a_\text{cav} = - \sum_i \frac{\partial a_\text{exp}}{\partial \bw_i} \delta\w_i.
\end{equation}
The shifts $\delta\w_i$ that appear on the rhs are functions of $\w_i$ and $g/2$, but can now be recast in terms of $\bw_i$ by using the zeroth-order relations in Eqs.~\eqref{eq:a_exp}, \eqref{eq:def_w}, and \eqref{eq:w_relations} once more. Throughout both stages, judicious use of the hierarchy in Eq.~\eqref{eq:hierarchy} was made to keep only the terms relevant at the level of the experimental precision. The end result is
\begin{align}
\label{eq:da_cav_full}
\delta a_\text{cav} =&\;
\frac{\beta_m}{\Mp \bw_c^2}\left( \frac{\phi_{rr}}{r_0^2} + \frac{\phi_{zz}}{2z_0^2} \right)
\nonumber\\
&- \frac{\beta_\gamma}{\Mp\bw_c^2} \left( \frac{\bw_a}{2 m_e}\frac{\phi_{rr}}{r_0^2} + \frac{49 \bw_z}{8 m_e} \frac{\phi_{zz}}{z_0^2} \right).
\end{align}

Note that the coefficient of $\phi_{zz}$ in the second line contains a factor of $2n_z + 1 = 49$. Notice also that the second line, arising from the classical vacuum polarization effect due to the photon coupling (Sec.~\ref{sec:cav_em}), is strongly suppressed by factors of
\begin{equation*}
\bw_a/m_e \sim \bw_z/m_e \sim 10^{-12}.
\end{equation*}
As a consequence, this effect is unable to place any meaningful constraint on the photon coupling. While we initially imagined that the large magnetic field in the cavity would be helpful for such a purpose, on the contrary, it turns out to offer little advantage because of the particular combination of frequencies that have to be measured. The correction $\delta\mathbf{A}$ couples to the orbital angular momentum while $\delta\mathbf{B}$  couples to the spin in the Hamiltonian [see Eq.~\eqref{eq:cav_dH}], and the two contributions approximately cancel out when computing $\delta a_\text{cav}$. The leading effect that survives is due to the correction $\delta V$ to the electrostatic potential. This is much smaller, since the ratio of the electric to magnetic energy densities is
\begin{equation}
\label{eq:cav_E/B}
\frac{\mathbf{E}^2}{\mathbf{B}^2} \sim \frac{V_0^2}{B_0^2 d^2} \sim 10^{-10}.
\end{equation}

Moving forward, we shall neglect any effect of the photon coupling on the cavity shift. Fortuitously, the combination of second derivatives in the first line of Eq.~\eqref{eq:da_cav_full} is exactly the Laplacian evaluated at the origin. Use of Eq.~\eqref{eq:eom_scalar} allows us to rewrite this in terms of $V_{\text{eff},\phi}$, such that
\begin{equation}
\label{eq:da_cav}
\delta a_\text{cav} = \frac{\beta_m(\phi_0) V_{\text{eff},\phi}(\phi_0)}{2\Mp \bw_c^2}.
\end{equation}
This effect contributes to the total deviation as $\delta a \supset - \delta a_\text{cav}$, where the minus sign can be traced back to the relative sign between $a_\text{SM}$ and $a_\text{exp}$ in Eq.~\eqref{eq:a_compare}.

\section{Chameleon constraints}
\label{sec:chm}

We have seen so far that a CLP generates additional quantum corrections and an experimental cavity shift that together contribute to a total deviation $\delta a$. This must be constrained according to Eq.~\eqref{eq:da_constraint} to respect the agreement between the Standard Model prediction and the experimental measurement of the electron's magnetic moment. Individual contributions to $\delta a$ are given in Eqs.~\eqref{eq:da_bMbM}, \eqref{eq:da_bMbG}, and \eqref{eq:da_cav}. In these equations, the calculations were carried out in complete generality, and the results are expressed in terms of the coupling strengths $\beta_m(\phi_0)$ and $\beta_\gamma(\phi_0)$, and the first derivative of the effective potential $V_{\text{eff},\phi}(\phi_0)$. Crucially, all three quantities depend only on the choice of model and the central field value $\phi_0$. To complete the calculation and determine the constraints on parameter space, we must simply specify the former and determine the latter. We do so for the chameleon in this section, and for the symmetron in the next.

The prototypical chameleon model assumes an inverse power-law potential\footnote{Note that the chameleon mechanism can also be realized with positive power-law potentials, $V(\phi) \propto \phi^{2s}$ with integer values of $s \geq 2$ \cite{Gubser:2004uf}, although we shall not consider such models in this work.} \cite{Khoury:2003aq,Ratra:1987rm,Wetterich:1987fm}
\begin{equation}
\label{eq:chm_potential}
V(\phi) = \frac{\Lambda^{4+n}}{\phi^n}
\quad (n>0)
\end{equation}
and coupling functions of the form
\begin{equation}
\Omega(\phi) = \exp\left( \frac{\phi}{M_c} \right),
\quad
\varepsilon(\phi) = \exp\left( \frac{\phi}{M_\gamma} \right).
\end{equation}
With these definitions, the dimensionless coupling strengths
\begin{equation}
\beta_m = \frac{\Mp}{M_c}, \quad \beta_\gamma = \frac{\Mp}{M_\gamma}
\end{equation}
are independent of the value of the field. Putting these together, the effective potential differentiates to
\begin{equation}
V_{\text{eff},\phi} = - \frac{n \Lambda^{4+n}}{\phi^{n+1}} + \left(\frac{\rho}{M_c} + \frac{\rho_\text{em}}{M_\gamma}\right).
\end{equation}

While, in principle, all of parameter space is open to exploration, focus has primarily been devoted to models in which $\Lambda$ is chosen to be near the dark energy scale, $\Lambda = 2.4\,\text{meV}$. This choice makes the chameleon cosmologically relevant, if we view the potential in Eq.~\eqref{eq:chm_potential} as just the leading $\phi$-dependent term in an expansion
\begin{equation}
V(\phi) = \Lambda^4 f(\Lambda^n/\phi^n) \simeq \Lambda^4 + \frac{\Lambda^{4+n}}{\phi^n},
\end{equation}
assumed to arise from nonperturbative effects \cite{Brax:2004qh}. The constant piece $\Lambda^4$ has no effect on laboratory scales, but is an alternative to $\Lambda$CDM for driving the accelerated expansion of the Universe. (See Refs.~\cite{Brax:2004qh,Wang:2012kj} for more on the cosmology of the chameleon.)

\subsection{Analytic estimates}
\label{sec:chm_1D}

As stated in Sec.~\ref{sec:cav_profile}, it is difficult to solve Eq.~\eqref{eq:eom_scalar}---either exactly or approximately---for the chameleon profile in the interior of the Penning trap. This is because the cavity radius and height are of the same size, so the problem is strictly two-dimensional. However, as we are interested only in the central field value $\phi_0$, it turns out that analyzing an analogous one-dimensional cavity suffices to capture the most salient features of the solution. We discuss this one-dimensional ``toy model'' first, before turning to a numerical solution of the cylindrical geometry proper in Sec.~\ref{sec:chm_numerics}.

The toy model in question is the following: Consider a plane-parallel cavity in the region $z \in [-l,l]$ surrounded by walls on either side extending to infinity. The density of matter is assumed to be piecewise constant, such that
\begin{equation*}
\rho =
\begin{cases}
\rho_\text{cav}  & z \in [-l,l],
\\
\rho_\text{wall} & \text{otherwise}.
\end{cases}
\end{equation*}
We shall neglect the electric field in the cavity, as its energy density is much smaller than that of the magnetic field; see Eq.~\eqref{eq:cav_E/B}. In doing so, the electromagnetic energy density is also piecewise constant,
\begin{equation*}
\rho_\text{em} \simeq
\begin{cases}
B_0^2/2 & z \in[-l,l],
\\
0 & \text{otherwise}.
\end{cases}
\end{equation*}
In this setup, Eq.~\eqref{eq:eom_scalar} then reduces to
\begin{equation}
\label{eq:chm_1D}
\frac{\text{d}^2\phi}{\text{d}z^2} = V_{\text{eff},\phi}.
\end{equation}

An exact solution to this equation is known \cite{Brax:2011hb,Ivanov:2012cb,Ivanov:2016rfs}, but only for $n\in\{1,2\}$ and when the interior of the cavity is pure vacuum. This is not general enough for our purposes. Instead, we use a standard technique to approximate the solution by solving linearized versions of Eq.~\eqref{eq:chm_1D} inside and outside the cavity, and imposing matching conditions at the adjoining boundaries \cite{Khoury:2003rn,Hinterbichler:2011ca,Tamaki:2008mf,Burrage:2014daa}. The linearized field equations are
\begin{equation}
\frac{\text{d}^2\phi}{\text{d}z^2} \simeq
\begin{cases}
m^2_0 (\phi-\phi_0) + V'_0 & |z| \leq l,
\\
m_\infty^2(\phi-\phi_\infty) & |z| > l.
\end{cases}
\end{equation}
In the interior of the cavity, we have expanded about the as-of-yet unknown central field value~$\phi_0$. The effective mass $m_0$ was defined previously in Eq.~\eqref{eq:def_mphi} as the second derivative $m_0^2 = V_{\text{eff},\phi\phi}(\phi_0)$ evaluated at the center. The constant term $V'_0 \coloneq V_{\text{eff},\phi}(\phi_0)$. Deep inside the walls, the chameleon will asymptote to the field value $\phi_\infty$ which minimizes the local effective potential, $V_{\text{eff},\phi}(\phi_\infty;\rho=\rho_\text{wall}) = 0$.
Solving this equation yields
\begin{equation}
\label{eq:chm_phi_infty}
\phi_\infty = \left( \frac{n \Lambda^{4+n} M_c}{\rho_\text{wall}} \right)^{1/(1+n)}.
\end{equation}
We have thus expanded the field equation in the walls about this point, with a mass $m_\infty^2 = V_{\text{eff},\phi\phi}(\phi_\infty)$ similarly defined.

Solving these equations brings about four integration constants, which are determined uniquely by the boundary conditions. Two of them are
\begin{equation}
\left.\frac{\text{d}\phi}{\text{d}z}\right|_{z=0} = 0,
\quad
\phi(z \to\pm\infty) = \phi_\infty,
\end{equation}
while the remaining two come from imposing continuity of $\phi(z)$ and its first derivative at $|z| = l$. With these considerations, the solution in the cavity ($|z| \leq l$) is
\begin{subequations}
\begin{equation}
\label{eq:chm_pp_int}
\phi(z) =
\phi_0 - \frac{V'_0}{m_0^2} - \frac{(\phi_0 - \phi_\infty - V_0'/m_0^2)\cosh(m_0 z)}{\cosh(m_0 l)+(m_0/m_\infty)\sinh(m_0 l)},
\end{equation}
whereas the solution in the walls ($|z| > l$) is
\begin{equation}
\label{eq:chm_pp_ext}
\phi(z) = \phi_\infty + \frac{(\phi_0 - \phi_\infty - V_0'/m_0^2) e^{-m_\infty(|z|-l)}}{1 + (m_\infty/m_0) \coth(m_0 l)}.
\end{equation}
\end{subequations}

An implicit equation for the central field value $\phi_0$ is obtained by demanding the solution in Eq.~\eqref{eq:chm_pp_int} satisfy the self-consistency condition
\begin{equation}
\label{eq:chm_solve_phi0}
\phi(z = 0) = \phi_0.
\end{equation}
Two approximations can be made to simplify this result. (Their implications and validity are discussed in the next two subsections.) First, let us assume that once in the walls, the chameleon quickly reaches its limiting value $\phi_\infty$. By inspecting Eq.~\eqref{eq:chm_pp_ext}, this will be true if $m_\infty \gg m_0$. Second, let us also assume that the interior of the cavity is pure vacuum, such that $V'_0 \simeq -n \Lambda^{4+n}/\phi_0^{n+1}$. When both these assumptions hold, Eq.~\eqref{eq:chm_solve_phi0} simplifies to
\begin{equation}
\label{eq:chm_1D_phi0}
\cosh(m_0 l) = n+2.
\end{equation}

This result admits an intuitive physical interpretation: In a vacuum cavity, the chameleon adjusts itself until its local Compton wavelength $m_0^{-1}$ is on the order of the size of the cavity~$l$ \cite{Khoury:2003rn}. This feature appears to be generic. A similar calculation can be found in Ref.~\cite{Brax:2007hi} for the case of an infinitely-long cylindrical cavity. The same result was obtained, except with the hyperbolic cosine replaced by the modified Bessel function of the first kind.

We expect this result to extend to higher dimensions also, although now the function appropriate to the geometry is not known. To proceed, we first note that Eq.~\eqref{eq:chm_1D_phi0} can be approximated by
\begin{equation}
m_0^2 l^2 \simeq 17.4 \frac{n+1.05}{n+10.5}
\end{equation}
for $n$ of order unity, where the rhs is the [1/1]-order Pad\'e approximant of $[\cosh^{-1}(n+2)]^2$ about $n=1$. For arbitrary (convex) cavity shapes, we conjecture that this generalizes to
\begin{equation}
\label{eq:chm_1D_pade}
m_0^2 l^2 \simeq \frac{n+1}{n+\delta},
\end{equation}
where $\delta$ is a constant depending on the geometry and any overall normalization of the rhs can been absorbed into the constant~$l$, which should now be thought of as a characteristic length scale of the cavity. Rearranging this equation and using the definition of $m_0$, we predict that the central field value has a dependence on $\Lambda$ and $n$ given by
\begin{equation}
\label{eq:chm_phi0}
\phi_0 \simeq \left[ n(n+\delta) \Lambda^{4+n} l^2 \right]^{1/(2+n)}.
\end{equation}
The two constants $(l,\delta)$ act as free parameters which should be tuned to best fit the numerical results.

\subsection{Numerical results}
\label{sec:chm_numerics}

\begin{figure}
\includegraphics[width=70mm]{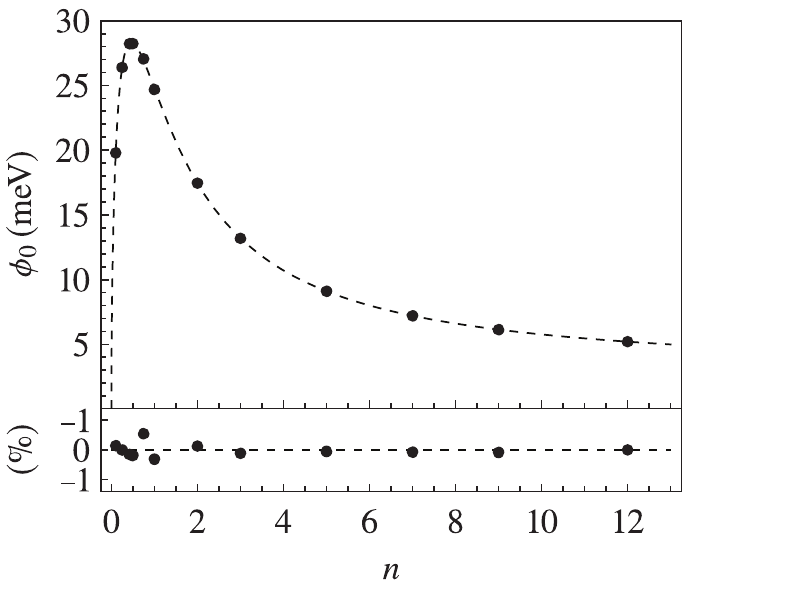}
\caption{Best-fitting analytic approximation (dashed line) to the central field value $\phi_0$ of the chameleon in the cylindrical vacuum cavity for different values of $n$ with $\Lambda = 2.4\,\text{meV}$, compared with the numerical results (black dots). The lower plot displays the percentage difference between the numerical and analytic results: All points agree to less than one percent.}
\label{fig:chm_phi0}
\end{figure}

\begin{figure}
\includegraphics[width=70mm]{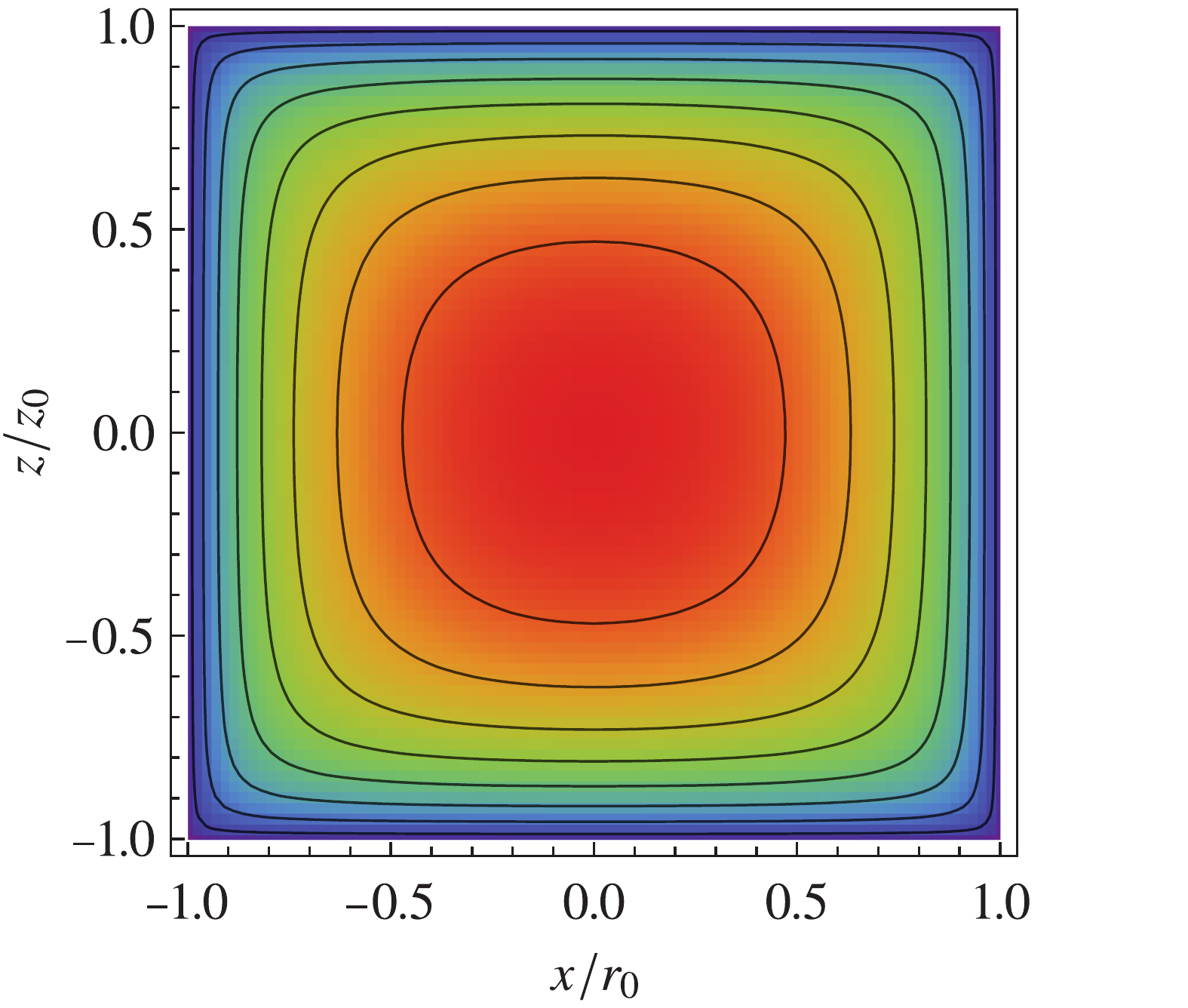}
\caption{Chameleon profile in the cylindrical vacuum cavity for $n=1$ and $\Lambda = 2.4\,\text{meV}$. The field value along the innermost contour is 90\% of the value at the origin. Moving outwards, successive contours are 80\%, 70\%, etc.~of the central field value. The field reaches 10\% near the boundary of the cavity, before quickly plummeting to $\phi\approx 0$ once inside the walls.}
\label{fig:chm_profile}
\end{figure}

We determine the full, nonlinear chameleon profile in the cylindrical Penning trap numerically by integrating Eq.~\eqref{eq:eom_scalar} through successive under-relaxation using the Gauss-Seidel scheme \cite{Press:2007:NRE:1403886} for 12 values of $n \in (0,13)$ with $\Lambda = 2.4\,\text{meV}$. Our code has been previously used to study similar problems in Ref.~\cite{Elder:2016yxm}, where more details on the method can be found. The dependence of $\phi_0$ on $n$ is shown in Fig.~\ref{fig:chm_phi0}, alongside the best-fitting analytic approximation, given in Eq.~\eqref{eq:chm_phi0}. The values of the best-fitting parameters are\footnote{These values, and analogous ones in Sec.~\ref{sec:sym}, were determined using the native \texttt{NonlinearModelFit} routine in \emph{Mathematica}.}
\begin{equation*}
l = 1.40\,\text{mm}, \quad
\delta = 2.78.
\end{equation*}
For illustrative purposes, we also present the full chameleon profile for $n=1$ in Fig.~\ref{fig:chm_profile}. The profiles for the remaining values of $n$ are qualitatively similar.

\begin{figure*}
\includegraphics[width=\textwidth]{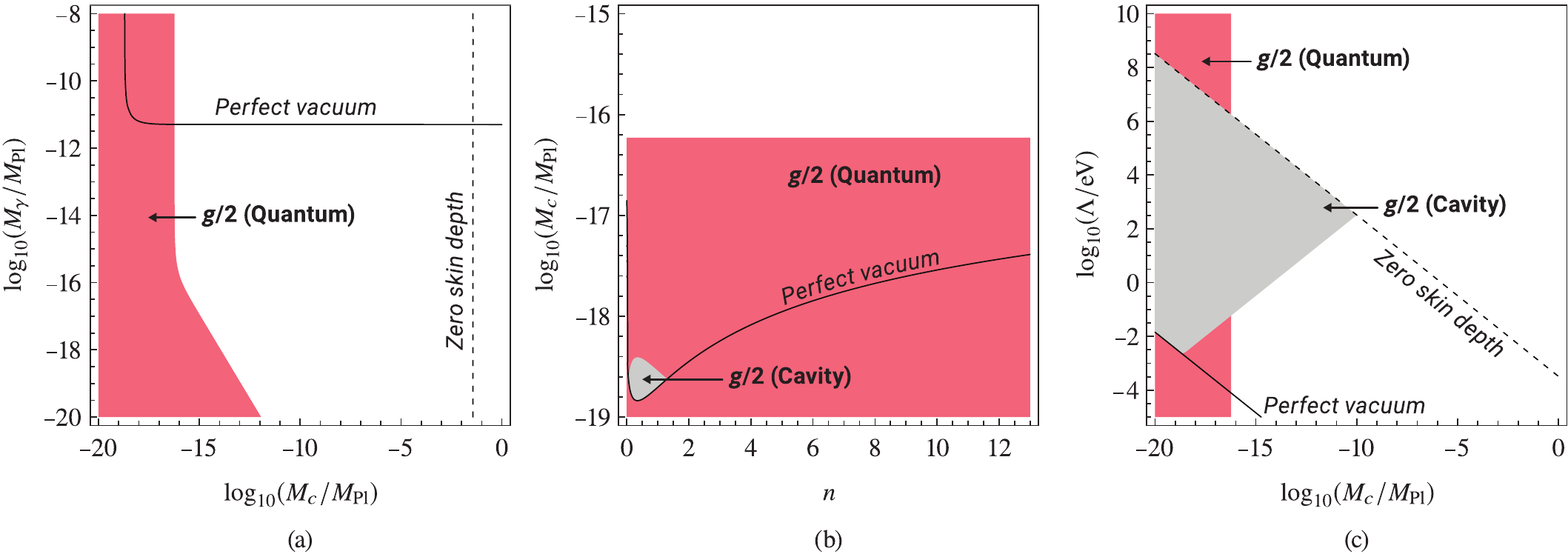}
\caption{Constraints on chameleon models due to the electron magnetic moment. The shaded regions are excluded at the 95\% confidence level. The panels correspond to the following slices in parameter space: (a) $n=1$, $\Lambda = 2.4\,\text{meV}$; (b) $\Lambda=2.4\,\text{meV}$, $\beta_\gamma \coloneq \Mp/M_\gamma = 0$; (c) $n=1$, $\beta_\gamma = 0$. Numerical limitations mean that the cavity shift can be computed reliably only when both the zero-skin-depth and perfect-vacuum approximations are valid (see Sec.~\ref{sec:chm_numerics} for details). This corresponds to the region above the solid line, and below or to the left of the dotted line. Inside this region, the constraints arising from the cavity shift are shaded in gray. Outside this region, only the constraints from the quantum corrections (pink), which are still reliable, are shown.
}
\label{fig:chm_c_1}
\end{figure*} 

Our approach is tractable only under two simplifying assumptions---the same as were made in the preceding subsection. We now give them names and discuss their implications:
\begin{enumerate}
\item\emph{Zero-skin-depth approximation}: We assume that the chameleon approaches its limiting value $\phi_\infty$ rapidly once inside the walls, such that we can approximate $\phi \approx \phi_\infty \approx 0$ at the boundary of the cavity. This is exactly true in the limit $\rho_\text{wall} \to \infty$, but will hold in practice provided
\begin{equation}
\label{eq:chm_approx_zsd}
m_0^2 \ll m_\infty^2.
\end{equation}
This approximation is essential, because in reality the walls of the cavity do not extend to infinity. By assuming that the chameleon quickly reaches $\phi_\infty$, we are assured that it has effectively decoupled itself from everything else happening beyond the walls, so that it is safe to neglect the complicated configuration of apparatuses surrounding the cavity.

\item\emph{Perfect-vacuum approximation}: We also assume that the interior of the cavity is a perfect vacuum. This is formally the limit $\rho_\text{cav}, \rho_\text{em} \to 0$, but will hold in practice provided
\begin{equation}
\label{eq:chm_approx_pv}
\frac{\rho_\text{cav}}{M_c} + \frac{\rho_\text{em}}{M_\gamma} \ll \frac{n \Lambda^{4+n}}{\phi_0^{n+1}}.
\end{equation}
This approximation is computationally convenient because the chameleon field equation reduces to
\begin{equation}
\nabla^2\phi = - \frac{n \Lambda^{4+n}}{\phi^{n+1}}
\end{equation}
in this limit. It is obvious that the central field value $\phi_0$ can then only depend on $\Lambda$ and $n$. More importantly, this equation admits the scaling symmetry
\begin{equation}
\label{eq:chm_lambda_scaling}
\Lambda \to f \Lambda, \quad
\phi \to f^{(4+n)/(2+n)} \phi,
\end{equation}
hence it suffices to perform the numerical integration for just one value of $\Lambda$; all other solutions are then accessible by rescaling.
\end{enumerate}

\subsection{Constraints}
\label{sec:chm_discussion}

The chameleon model contains four free parameters $(n,\Lambda,M_c,M_\gamma)$ which we wish to constrain. In terms of these parameters, the total deviation $\delta a$ takes the form
\begin{equation}
\label{eq:chm_da}
\delta a =
\frac{1}{2 M_c \bw_c^2}\frac{n \Lambda^{4+n}}{\phi_0^{n+1}}
+
3\left(\frac{m_e}{4\pi M_c}\right)^2
+
\frac{10}{M_c M_\gamma} \left(\frac{m_e}{4\pi}\right)^2,
\end{equation}
where the first term is due to the cavity shift, while the remaining two arise from the quantum corrections.

The cavity shift term exhibits a strong dependence on the central field value $\phi_0$, which we can predict reliably using Eq.~\eqref{eq:chm_phi0} only when both the zero-skin-depth (ZSD) and perfect-vacuum (PV) approximations are valid. As the limit $\rho \to 0$ is equivalent to taking $M_c,M_\gamma \to \infty$, these approximations are easily satisfied in some regions of parameter space, but break down in others. (The boundary at which this happens is estimated in Appendix~\ref{app:approx}.) For easy reference, we shall refer to the region where both the ZSD and PV approximations hold as the numerically accessible region (NAR). Outside this NAR, we no longer have a good sense for how $\phi_0$ behaves, and consequently cannot determine constraints arising from the cavity shift. In contrast, the quantum correction terms extend well beyond the NAR, since this effect has virtually no dependence on $\phi_0$ as long as $m_0 \ll m_e$. (The boundary at which this approximation breaks down is also discussed in Appendix~\ref{app:approx}.) Regions of parameter space excluded at the 95\% confidence level by the cavity shift and quantum corrections are shown, separately, in Fig.~\ref{fig:chm_c_1}.

For $n=1$ and $\Lambda = 2.4\,\text{meV}$, the chameleon field profile near the center is sufficiently flat that the cavity shift has no impact within the NAR. The constraints in Fig.~\ref{fig:chm_c_1}(a) are thus set entirely by the quantum corrections. Note that the effect of the photon coupling only becomes noticeable for $\log_{10}(M_\gamma/\Mp) \lesssim -16$, although couplings in the region $\lesssim -15.4$ are already ruled out from considering collider experiments \cite{Brax:2009aw}. We therefore find that the electron's magnetic moment places no meaningful constraint on the photon coupling scale $M_\gamma$. This statement is true for all values of $\Lambda$ and $n$, since the quantum corrections are independent of these parameters, at least at leading one-loop order.

Focusing on the matter coupling scale $M_c$ now, the quantum corrections provide a universal lower bound
\begin{equation*}
\log_{10}(M_c/\Mp) \gtrsim -16.7
\end{equation*}
independent of $\Lambda$ and $n$, as shown in Fig.~\ref{fig:chm_c_1}(b). This is a weak constraint, stemming from the small ratio $(m_e/M_c)^2 \ll 1$ that sets the scale of the quantum corrections. Other experiments do much better. Most notably, a different precision QED test---measurement of the $1S$--$2S$ transition in hydrogen---gives a slightly better lower bound $\log_{10}(M_c/\Mp) \gtrsim -14$ \cite{Brax:2010gp,Wong:2017jer}. The best lower bound to date, however, comes from atom interferometry \cite{Burrage:2014oza,Burrage:2015lya,Brax:2016wjk,Hamilton:2015zga,Jaffe:2016fsh,Elder:2016yxm}. Depending on the value of $n$, the lower bound is between $-4$ to just under $-2.5$.

\begin{figure}
\includegraphics[width=68mm]{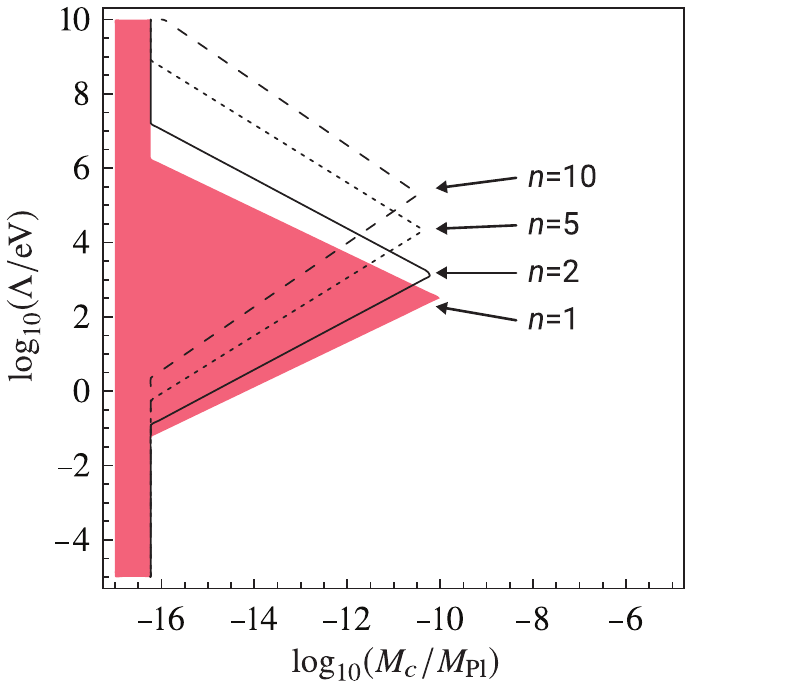}
\caption{Constraints on the chameleon due to the electron magnetic moment in the $M_c$--$\Lambda$ plane. Parameters in the shaded region are excluded for the $n=1$ chameleon at the 95\% confidence level. The regions to the left of the solid, dotted, and dashed lines rule out parameters for other illustrative values of $n$.}
\label{fig:chm_c_3}
\end{figure}

\begin{figure}
\includegraphics[width=68mm]{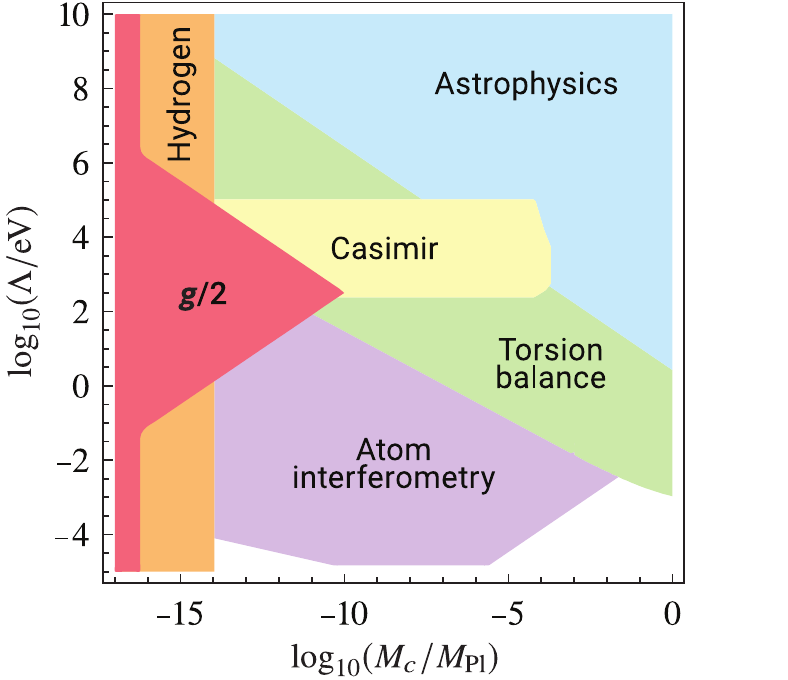}
\caption{The constraining power of the electron magnetic moment for the $n=1$ chameleon, compared with a selection of other experiments \cite{Brax:2007vm,Brax:2010gp,Upadhye:2012qu,Jain:2012tn,Vikram:2014uza,Jaffe:2016fsh}. See Ref.~\cite{Burrage:2017qrf} for details on all existing constraints.}
\label{fig:chm_c_2}
\end{figure}

Moving away from the dark energy scale, increasing $\Lambda$ drives the chameleon to climb to a larger central field value. When this happens, the cavity shift dominates until $\Lambda$ becomes too large, at which point we impinge on the boundary of the NAR. The end result is a triangular-shaped region excluded by this effect, as shown in Fig.~\ref{fig:chm_c_1}(c). For $n=1$, the lower bound on $M_c$ extends all the way out to $\log_{10}(M_c/\Mp) = -10$ when $\Lambda \approx 300\,\text{eV}$. The shape of the excluded region is qualitatively similar for other values of $n$, as shown in Fig.~\ref{fig:chm_c_3}.

A comparison of our constraints with those from a selection of other experiments is shown in Fig.~\ref{fig:chm_c_2} for the $n=1$ chameleon. Although we do not cover any new region of parameter space not already ruled out by other experiments, it is worth remarking that our results represent the tightest constraints yet achievable by an experiment not originally designed to search for fifth forces.

\section{Symmetron constraints}
\label{sec:sym}

The symmetron model is characterized by a Higgs-like, double-well potential
\begin{equation}
V(\phi) = -\frac{1}{2} \mu^2 \phi^2 + \frac{\lambda}{4} \phi^4
\end{equation}
and coupling functions
\begin{align}
\Omega(\phi) &= 1 + \frac{\phi^2}{2 M_s^2} + \mathcal O\left(\frac{\phi^4}{M_s^4}\right),
\nonumber\\
\varepsilon(\phi) &= 1 + \frac{\phi^2}{2 M_\gamma^2} + \mathcal O\left(\frac{\phi^4}{M_\gamma^4}\right)
\end{align}
consistent with the field's $\phi \to -\phi$ symmetry. Differentiation gives the field-dependent dimensionless coupling strengths
\begin{equation}
\beta_m(\phi) = \Mp \frac{\phi}{M_s^2}, \quad
\beta_\gamma(\phi) = \Mp \frac{\phi}{M_\gamma^2}
\end{equation}
to leading order. Taken altogether, these yield an effective potential
\begin{equation}
V_\text{eff}(\phi) = \frac{1}{2}\mu^2\left( \frac{\rho}{\mu^2 M_s^2} + \frac{\rho_\text{em}}{\mu^2 M_\gamma^2} - 1 \right) \phi^2 + \frac{\lambda}{4} \phi^4.
\end{equation}

\subsection{Analytic estimates}
\label{sec:sym_1D}

As we did for the chameleon, it is helpful to first consider an analogous plane-parallel cavity whose solution will elucidate the relevant physics. Unlike the chameleon, this simple toy model admits an exact solution even in the presence of matter, provided only that it is distributed in a piecewise-constant fashion. The only spatially-varying source of matter is the energy density in the electric field, which for all intents and purposes is small enough to be neglected [recall Eq.~\eqref{eq:cav_E/B}]. Doing so, the symmetron's field equation can be integrated up once to give
\begin{equation}
\left(\frac{\text{d}\phi}{\text{d}z}\right)^2 = \frac{\mu^2}{2}\left( \frac{\rho}{\mu^2 M_s^2} + \frac{\rho_\text{em}}{\mu^2 M_\gamma^2} - 1 \right) \phi^2 + \frac{\lambda}{4} \phi^4 + \text{const.},
\end{equation}
with the constant determined by boundary conditions.

Inside the cavity, let us define an effective mass scale
\begin{equation}
\mu_0^2 = \mu^2\left( 1 - \frac{\rho_\text{cav}}{\mu^2 M_s^2} - \frac{\rho_\text{em}}{\mu^2 M_\gamma^2} \right),
\end{equation}
which must satisfy $\mu_0^2 > 0$ as a necessary condition if the symmetron is to break its $\mathbb Z_2$ symmetry. When this is the case, we expect the field to climb to an as-of-yet unknown value $\phi_0$ in the center, assumed to be a local maximum satisfying
\begin{equation}
\label{eq:sym_1D_bc_grad}
\left.\frac{\text{d}\phi}{\text{d}z}\right|_{z=0} = 0.
\end{equation}
Indeed, if the cavity were infinitely large, the field would have sufficient room to minimize its effective potential, such that $\phi_0 \to \pm \mu_0/\sqrt\lambda$. As this gives the largest possible value for~$|\phi_0|$, it is convenient to define a dimensionless scalar field
\begin{equation}
\varphi = \frac{\phi}{\mu_0/\sqrt\lambda}
\end{equation}
with range $\varphi \in [-1,1]$. With this definition, the symmetron field equation inside the cavity ($|z| \leq l$) becomes
\begin{equation}
\frac{1}{\mu_0^2} \left(\frac{\text{d}\varphi}{\text{d}z}\right)^2
=
- (\varphi^2-\varphi_0^2) + \frac{1}{2}(\varphi^4-\varphi_0^4),
\end{equation}
which crucially depends only on the parameter $\mu_0$. This first-order differential equation can be integrated to yield \cite{Upadhye:2012rc}
\begin{equation}
- \mu_0 z \frac{\varphi_0}{\sqrt{2v^2}} =
F\left(\sin^{-1}\left(\frac{\varphi(z)}{\varphi_0}\right), v \right)
- K(v),
\end{equation}
where we have chosen the positive branch $\varphi(z) > 0$ without loss of generality, and have defined $v^2 = \varphi_0^2/(2-\varphi_0^2)$. This result is expressed in terms of the elliptic integrals of the first kind
\begin{equation}
F(u,v) = \int_0^u \frac{\text{d}\theta}{\sqrt{1- v^2 \sin^2\theta}},
\end{equation}
and $K(v) = F(\pi/2,v)$. From the definitions of the Jacobi elliptic functions
\begin{align}
\text{sn}(u,v) &= \sin F^{-1}(u,v),
\nonumber\\
\text{cn}(u,v) &= \cos F^{-1}(u,v),
\nonumber\\
\text{dn}(u,v) &= \sqrt{1- v^2 \text{sn}^2(u,v)},
\end{align}
this can be inverted to give
\begin{equation}
\varphi(z) = \varphi_0\,\text{sn}\left( - \mu_0 z \frac{\varphi_0}{\sqrt{2v^2}} + K(v),v\right).
\end{equation}
As a final step, note that the elliptic functions satisfy the identity
\begin{equation}
\text{sn}(u + K(v),v) = \frac{\text{cn}(u,v)}{\text{dn}(u,v)} \eqcolon \text{cd}(u,v),
\end{equation}
where the function $\text{cd}$ is even in its first argument. Hence, the exact solution for the symmetron field in the cavity is (see also Ref.~\cite{Brax:2017hna})
\begin{equation}
\label{eq:sym_1D_sol_int}
\varphi(z) = \varphi_0\,\text{cd}\left( \mu_0 z \frac{\varphi_0}{\sqrt{2v^2}}, v \right).
\end{equation}

\begin{figure*}
\includegraphics[width=\textwidth]{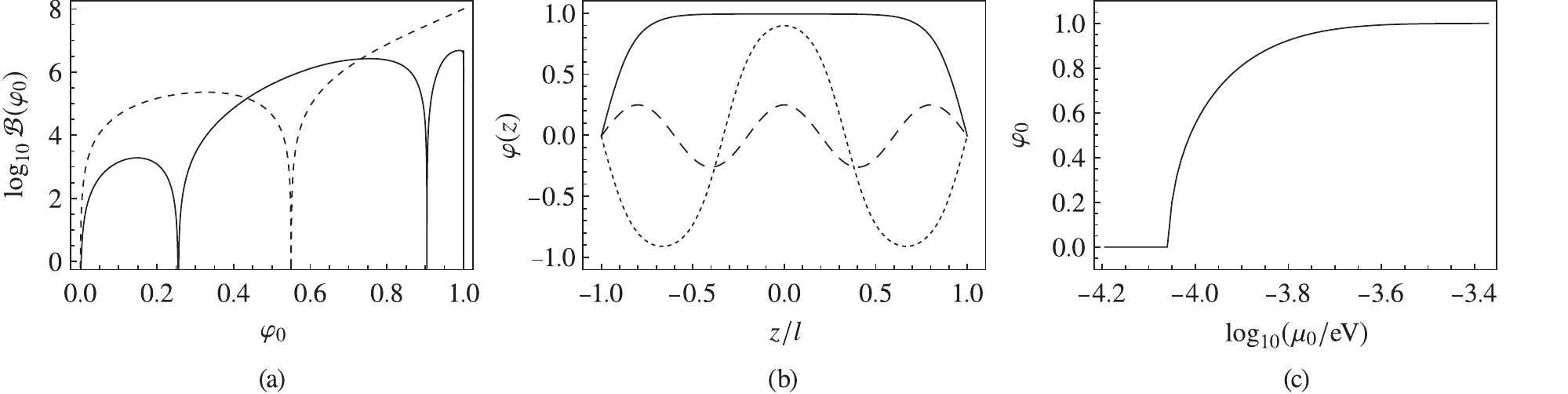}
\caption{(a) The central field value $\varphi_0$ of the symmetron in a plane-parallel cavity is determined by finding the root(s) of the function $\mathcal B(\varphi_0)$, shown for two illustrative values $\mu_0 = 0.1\,\text{meV}$ (dashed line) and $0.45\,\text{meV}$ (solid line). (b) Symmetron profiles corresponding to the roots $\varphi_0 \approx \{0.26, 0.90, 1.00 \}$ for $\mu_0 = 0.45\,\text{meV}$ are shown as dashed, dotted, and solid lines, respectively. (c) The largest root $\varphi_0$ as a function of the mass scale $\mu_0$. In all three panels, illustrative values $\mu_\infty = 1\,\text{eV}$ and $l = 3.5\,\text{mm}$ are used.}
\label{fig:sym_1D}
\end{figure*}

Similarly, the solution in the walls ($|z| \geq l$) is governed by the equation
\begin{equation}
\label{eq:sym_1D_ode_walls}
\frac{1}{\mu_0^2} \left(\frac{\text{d}\varphi}{\text{d}z}\right)^2
=
\left(\frac{\mu_\infty}{\mu_0}\right)^2 \varphi^2 + \frac{1}{2}\varphi^4,
\end{equation}
made to satisfy the boundary condition $\varphi(|z| \to \infty) = 0$. The corresponding effective mass scale $\mu_\infty$ is defined by
\begin{equation}
\mu_\infty^2 = \mu^2 \left(\frac{\rho_\text{wall}}{\mu^2 M_s^2} - 1 \right),
\end{equation}
which must be positive to restore the $\mathbb Z_2$ symmetry in this region.

If we were so inclined, Eq.~\eqref{eq:sym_1D_ode_walls} can then be integrated to give the exact solution in the walls, with the integration constant determined by requiring continuity of $\varphi$ at the boundary $|z| = l$. A self-consistency equation for $\varphi_0$ is then obtained by also demanding continuity of the first derivatives. However, as we are here only interested in the solution within the cavity, this process can be sidestepped in favor of a shortcut. An equivalent self-consistency condition can be obtained by substituting Eq.~\eqref{eq:sym_1D_sol_int} into Eq.~\eqref{eq:sym_1D_ode_walls} evaluated at $|z| = l$. This yields an implicit equation for the central field value $\varphi_0$.

Said again in different words, we  can solve for $\varphi_0$ by searching for the root of the function
\begin{equation}
\label{eq:sym_1D_B}
\mathcal B(\varphi_0;\mu_0,\mu_\infty,l) = \left.\left(\frac{\mu_\infty}{\mu_0}\right)^2 \varphi^2 + \frac{\varphi^4}{2} - \frac{1}{\mu_0^2} \left(\frac{\text{d}\varphi}{\text{d}z}\right)^2 \right|_{z=l},
\end{equation}
where $\varphi(z)$ on the rhs is given by Eq.~\eqref{eq:sym_1D_sol_int}. The function $\mathcal B(\varphi_0)$ is drawn for two illustrative values of $\mu_0$ in Fig.~\ref{fig:sym_1D}(a). Above a certain threshold value of $\mu_0$, the function begins to admit multiple roots. Each root is a valid solution of the field equation, with smaller values of $\varphi_0$ corresponding to field configurations with an increasing number of nodes, as seen in Fig.~\ref{fig:sym_1D}(b). For an intuitive picture, we should view a symmetron bubble as a solitonic object of a certain minimum width specified by $\mu_0$. If the length scale set by this mass matches the size of the cavity, $\mu_0 l \sim \mathcal O(1)$, then a single bubble can be contained within the walls. For larger values of $\mu_0$, the characteristic size of each solitonic packet decreases, and thus it becomes possible to fit multiple nodes within the same available space. In fact, when this is the case, we can relax the boundary condition in Eq.~\eqref{eq:sym_1D_bc_grad} to also allow for odd solutions in the cavity. Such solutions are discussed further in Ref.~\cite{Brax:2017hna}.

In an experimental setup, however, it is natural to expect that the symmetron will occupy the state of lowest free energy, corresponding to the solution with only one antinode. This is given by the largest root $\varphi_0$; which is shown as a function of $\mu_0$ in Fig.~\ref{fig:sym_1D}(c). This curve is also easy to understand intuitively: For very small values of $\mu_0$, the symmetron has too large a Compton wavelength and is unable to resolve the size of the cavity, thus remains in its symmetry-unbroken phase, $\varphi_0 = 0$. At a threshold value of $\mu_0 l \sim 1.6$, the field is finally able to support a bubble that can fit within the cavity, and the curve starts to grow. For larger values of $\mu_0$, the curve starts its plateau at $\varphi_0 \approx 1$ when the Compton wavelength is sufficiently small that the field almost immediately reaches the minimum of its effective potential once inside the cavity. This qualitative picture holds also when we generalize to the two-dimensional cylindrical case in the next subsection.

\subsection{Numerical results}

\begin{figure}[b]
\includegraphics[width=75mm]{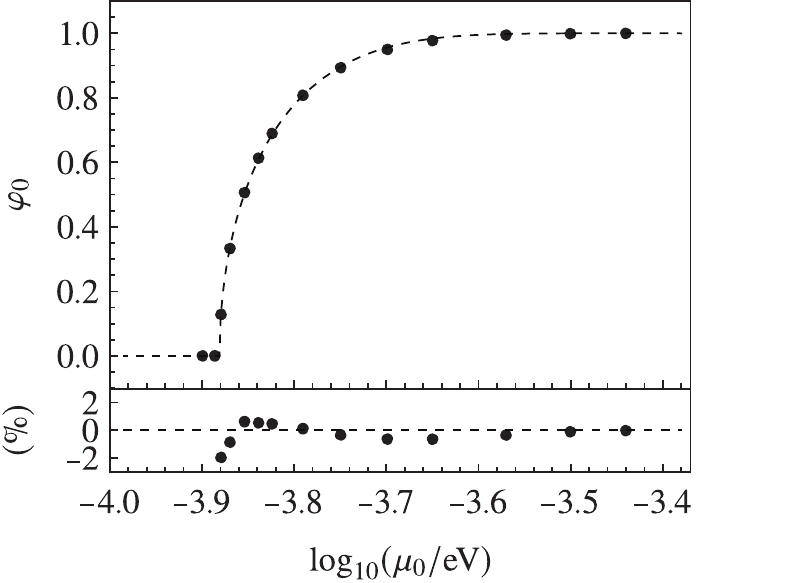}
\caption{Best-fitting analytic approximation (dashed line) to the dimensionless central field value $\varphi_0$ of the symmetron in the cylindrical vacuum cavity for different values of $\mu_0$, compared with the numerical results (black dots). The lower plot displays the percentage difference between the numerical and analytic results: All points agree to less than one percent, except the first three near $\mu_0 = 10^{-3.9}\,\text{eV}$ where $\varphi_0$ differs from zero only in the eighth (or higher) decimal place. Any discrepancy here is of no concern, since the numerical accuracy is unreliable for such small values of the field.}
\label{fig:sym_phi0}
\end{figure}

\begin{figure}[b]
\includegraphics[width=70mm]{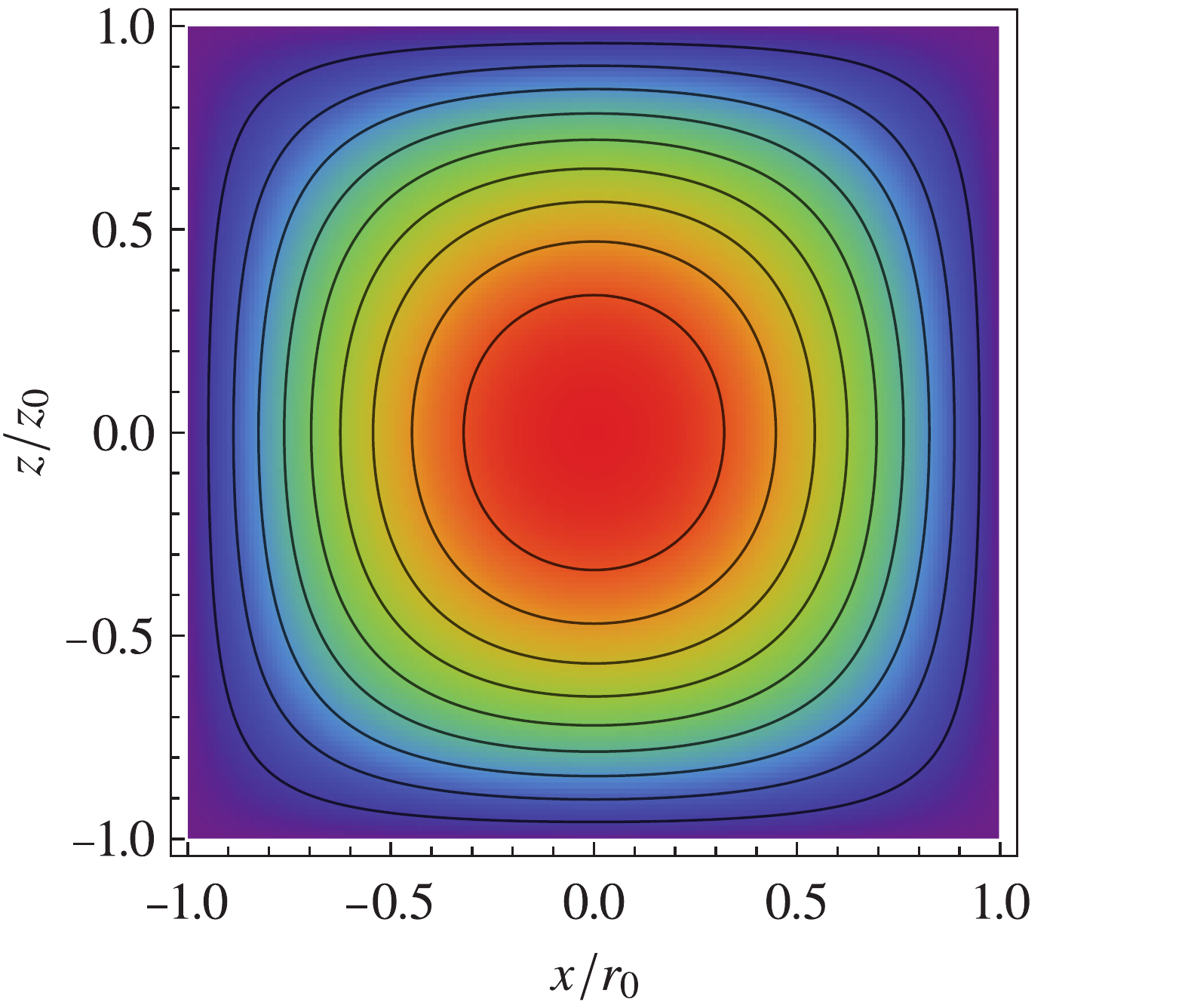}
\caption{Symmetron profile in the cylindrical vacuum cavity for $\mu_0 = 10^{-3.82}\,\text{eV}$. The field value along the innermost contour is 90\% of the value at the origin. Moving outwards, successive contours are 80\%, 70\%, etc.~of the central field value. The field reaches $\phi = 0$ once at the walls.}
\label{fig:sym_profile}
\end{figure}

The same numerical scheme as discussed in Sec.~\ref{sec:chm_numerics} is used to solve for the symmetron profile inside the cylindrical vacuum cavity. As we saw earlier, for this model the presence of piecewise-constant distributions of matter can be accounted for exactly by defining effective mass scales $\mu_0$ and $\mu_\infty$, hence only the zero-skin-depth (ZSD) approximation is needed. To recap, this assumes that the symmetron rapidly reaches its limiting value $\phi = 0$ once inside the walls, such that the field is essentially decoupled from its greater surroundings. Formally this is the limit $\rho_\text{wall}$ or $\mu_\infty \to \infty$, but will hold in practice provided
\begin{equation}
\label{eq:sym_zsd}
\mu_0^2 \ll \mu_\infty^2.
\end{equation}

\begin{figure*}
\includegraphics[width=\textwidth]{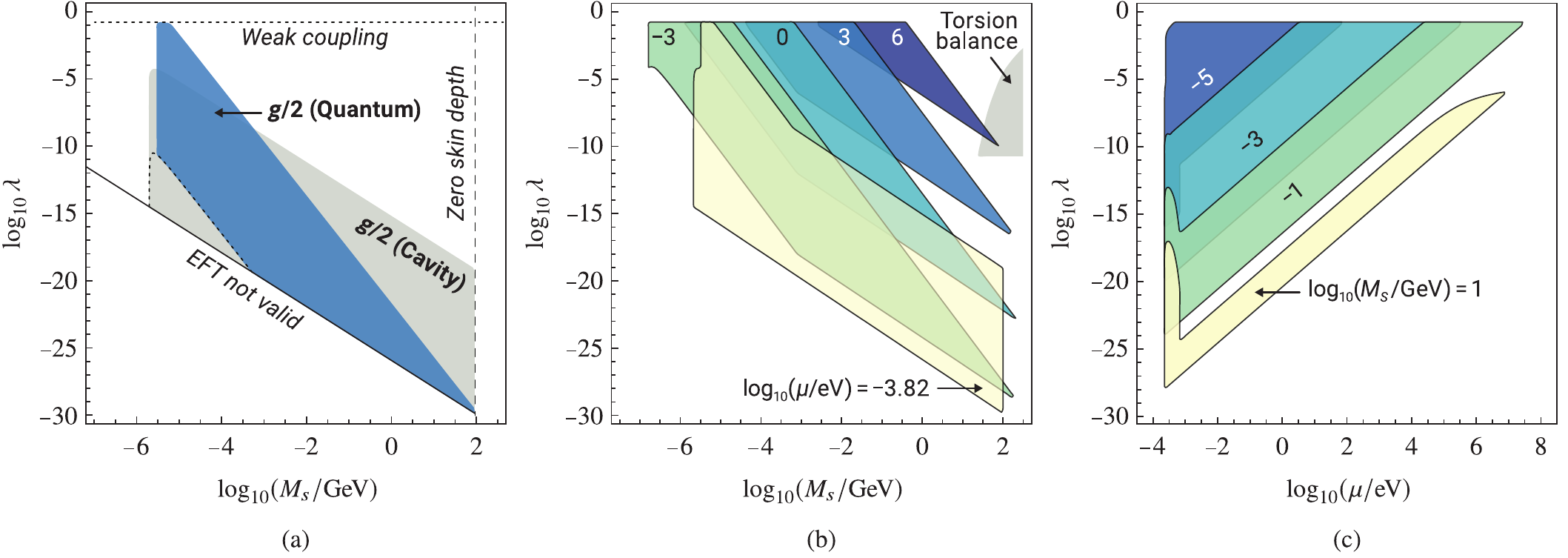}
\caption{Constraints on symmetron models due to the electron magnetic moment in the limit of a negligible photon coupling $M_\gamma \to \infty$. The shaded regions are excluded at the 95\% confidence level. Constraints arising from the cavity shift (gray) and quantum corrections (blue) are shown separately for the case $\mu = 10^{-3.82}\,\text{eV}$ in (a). Numerical limitations mean that these constraints can be computed reliably only when the zero-skin-depth (ZSD) approximation is valid, which explains the sharp cutoff for large $M_s$, as indicated by the vertical dashed line. Furthermore, the quantum correction terms are valid only in the weak coupling regime, corresponding to the region sandwiched between the dotted lines. Finally, no constraints are given for sufficiently small values of $\lambda$ when the EFT itself becomes unworkable, as shown by the solid line (see text in Sec.~\ref{sec:sym_c_matter} for more details). The combined constraints from the cavity shift and quantum corrections are shown together as one shaded region in (b) and (c) for different values of $\mu$. The same limits from assuming the ZSD approximation, weak coupling, and a valid EFT apply to each shaded region. For comparison, the region ruled out by torsion balance experiments \cite{Upadhye:2012rc} for $\mu = 10^{-3}\,\text{eV}$ is also shown in (b). In (c), observe that the parameter space is unconstrained for $\mu < 10^{-3.88}\,\text{eV}$, which is when the symmetron remains in its symmetry-unbroken phase inside the cavity. This same effect is responsible for the sharp cutoff at low $M_s$ in (a) and (b).}
\label{fig:sym_c_1}
\end{figure*}

We have performed the numerical integration for 15 values of $\mu_0$ in the range $\log_{10}(\mu_0/\text{eV}) \in (-4,-3)$, with the results of the dimensionless central field value $\varphi_0$ shown in Fig.~\ref{fig:sym_phi0}. The curve has a similar shape to what we found in the one-dimensional case, beginning its rise above zero at $\mu_0 \sim 10^{-3.88}\,\text{eV}$ and reaching the plateau by $\mu_0 \sim 10^{-3.39}\,\text{eV}$. For illustrative purposes, the full symmetron profile for the intermediate value $\mu_0 = 10^{-3.82}\,\text{eV}$ is shown in Fig.~\ref{fig:sym_profile}.

With some educated guessing, we have found that the curve in Fig.~\ref{fig:sym_phi0} can be well described by an empirical formula. Our starting point is the function $\mathcal B(\varphi_0)$ in Eq.~\eqref{eq:sym_1D_B}, the roots of which give the correct value of $\varphi_0$ in the one-dimensional case. Imposing the ZSD approximation, the limit $\mu_\infty \to \infty$ reduces this to the problem of finding the root of
\begin{equation}
\label{eq:sym_1D_approx}
\varphi(z=l) = \varphi_0\,\text{cd}\left( \mu_0 l \frac{\varphi_0}{\sqrt{2v^2}}, v \right) = 0.
\end{equation}
Finally, we introduce an \emph{ad hoc} parameter $\delta$ that deforms the solution away from the plane-parallel geometry, such that the new implicit equation for $\varphi_0$ is
\begin{equation}
\label{eq:sym_varphi0}
\varphi_0\,\text{cd}\left( (\mu_0 l)^{1+\delta} \frac{\varphi_0}{\sqrt{2v^2}}, v \right) = 0.
\end{equation}
This is given in terms of two free parameters $(l,\delta)$ which we should fit to the numerical data. Roughly speaking, the role of the characteristic length scale~$l$ is to fix the point at which the curve starts to rise above zero. The deformation parameter $\delta$ then tells us how quickly the curve reaches its plateau. The best-fitting parameters for the cylindrical Penning trap considered here are
\begin{equation*}
l = 1.96\,\text{mm}, \quad
\delta = 0.70.
\end{equation*}

\subsection{Constraints}

The symmetron model is specified by four parameters $(\mu,\lambda, M_s, M_\gamma)$ which we now constrain. In terms of these parameters, the total deviation $\delta a$ takes the form
\begin{align}
\label{eq:sym_da}
\delta a = &\,
\frac{\mu_0^4 \varphi_0^2(1-\varphi_0^2)}{2 \bw_c^2 M_s^2 \lambda}
+
\left(\frac{m_e}{4\pi}\right)^2 \frac{2\mu_0^2 \varphi_0^2}{M_s^4 \lambda} I_1(m_0/m_e)
\nonumber\\
&+
\left(\frac{m_e}{4\pi}\right)^2\frac{4\mu_0^2 \varphi_0^2}{M_s^2 M_\gamma^2 \lambda}
[1 +  I_2(m_0/m_e)],
\end{align}
where the first term is due to the cavity shift, while the remaining two arise from the quantum corrections. Unlike the chameleon which always satisfies $m_0/m_e \ll 1$, the effective symmetron mass in the cavity
\begin{equation}
m_0^2 = V_{\text{eff},\phi\phi}(\phi_0) = \mu_0^2(3\varphi_0^2 -1)
\end{equation}
can be made arbitrarily large by increasing the value of $\mu$. For this reason, we have retained the integrals $I_1$ and $I_2$ in Eq.~\eqref{eq:sym_da}.

\subsubsection{Matter coupling only}
\label{sec:sym_c_matter}

It is instructive to first neglect the photon coupling and focus on the subspace $(\mu,\lambda, M_s)$. Regions excluded at the 95\% confidence level are shown in Fig.~\ref{fig:sym_c_1}. Notice that the cavity shift term in Eq.~\eqref{eq:sym_da} is proportional to $\varphi_0^2(1-\varphi_0^2)$, thus switches off when $\varphi_0 = 0$ or $\varphi_0 = 1$. In terms of the symmetron mass, this means that the cavity shift exerts an appreciable force only in the small range $\mu \in [10^{-3.88}, 10^{-3.39}]\,\text{eV}$ (see Fig.~\ref{fig:sym_phi0}).\footnote{In most of the parameter space probed by this experiment, the mass scales $\mu$ and $\mu_0$ are essentially equivalent, and will be used interchangeably.} In Fig.~\ref{fig:sym_c_1}(a), constraints are shown for the illustrative value $\mu = 10^{-3.82}\,\text{eV} = 0.15\,\text{meV}$, which we have specifically chosen because it maximizes the quantity $\varphi_0^2(1-\varphi_0^2)$, and thus (approximately) maximizes the size of the cavity shift.

This sensitive dependence on $\mu$ is the reason why other laboratory experiments hitherto have left the symmetron parameter space mostly unexplored. Atom interferometry experiments \cite{Burrage:2016rkv,Jaffe:2016fsh}, for instance, place meaningful bounds only in the range $\mu \in [10^{-5}, 10^{-4}]\,\text{eV}$, whereas an analysis of torsion pendula \cite{Upadhye:2012rc} has so far only considered the range $[10^{-4},10^{-2}]\,\text{eV}$. This does not present an obstacle for the electron magnetic moment experiment, however, because in addition to the cavity shift, there exists also quantum correction terms that survive up to much larger values of $\mu$, which are primarily responsible for the constraints in Figs.~\ref{fig:sym_c_1}(b) and \ref{fig:sym_c_1}(c).

Having said that, not all of parameter space is accessible to this experiment. As always with the symmetron, the parameter space is unconstrained when spontaneous symmetry breaking fails to occur inside the cavity. This is the case for all values of $(\lambda, M_s,M_\gamma)$ when $\mu < 10^{-3.88}\,\text{eV}$. For larger masses, symmetry breaking occurs only above a minimum value of $M_s$, which explains the sharp cutoff at low $M_s$ seen in Figs.~\ref{fig:sym_c_1}(a) and \ref{fig:sym_c_1}(b). At the other end, the ZSD approximation breaks down beyond a maximum value of $M_s$---shown by the right vertical dashed line---at which point the central field value $\varphi_0$ can no longer be reliably predicted from Eq.~\eqref{eq:sym_varphi0}. Since every term in Eq.~\eqref{eq:sym_da} depends strongly on $\varphi_0$, constraints cannot be reliably determined to the right of this boundary. (Appendix~\ref{app:approx} describes how this boundary is estimated.)

Further limitations must be taken into account when determining the constraints arising from the quantum corrections. First, our perturbative approach requires a weak self-coupling\footnote{Recall that $\lambda$ appears in the potential as $V(\phi) \supset \lambda \phi^4/4$. However, when computing Feynman diagrams, the combinatorial factors are simplest if we organize the perturbative expansion in powers of $\lambda'$, where $\lambda'/4! = \lambda/4$. Imposing the condition $\lambda' \lesssim 1$ explains the factor of $1/6$ in Eq.~\eqref{eq:sym_validity_wc_1}.}
\begin{subequations}
\label{eq:sym_validity}
\begin{equation}
\label{eq:sym_validity_wc_1}
\lambda \lesssim 1/6.
\end{equation}
For the same reason, the Yukawa-like, scalar-matter coupling must also be weak [cf. Eq.~\eqref{eq:L_linear_couplings}],
\begin{equation}
\label{eq:sym_validity_wc_2}
\frac{\beta_m(\phi_0) m_e}{\Mp} = \frac{\mu_0^2 \varphi_0 m_e}{\sqrt\lambda M_s^2} \lesssim 1.
\end{equation}
For sufficiently small values of $\lambda$, the EFT itself becomes unworkable. For a rough estimate of when this happens, we shall deem it a necessary condition that the functions $\Omega(\phi)$ and $\varepsilon(\phi)$ do not deviate too far from unity. Inside the cavity, the (classical) symmetron field reaches a maximum value of at most $\phi_0 = \mu/\sqrt\lambda$, so our condition is satisfied provided
\begin{equation}
\label{eq:sym_validity_eft}
\frac{\mu^2}{2\lambda M_s^2} \lesssim 1, \quad
\frac{\mu^2}{2\lambda M_\gamma^2} \lesssim 1.
\end{equation}
\end{subequations}
The boundary lines demarcating the regions in parameter space that satisfy these conditions are shown in Fig.~\ref{fig:sym_c_1}(a). To prevent an overcrowded plot, they are not drawn again in Figs.~\ref{fig:sym_c_1}(b) and \ref{fig:sym_c_1}(c), nor in the remaining figures that follow, although it should be understood that they continue to be in effect.

\begin{figure*}
\includegraphics[width=\textwidth]{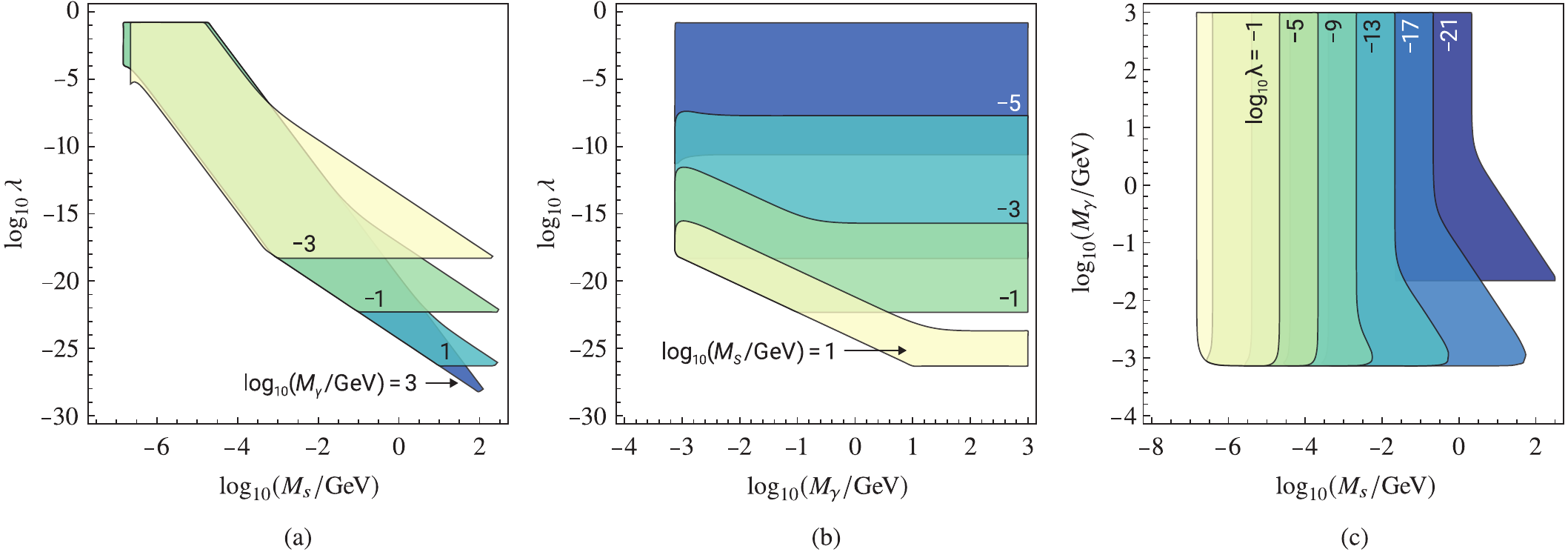}
\caption{Constraints on the $\mu = 10^{-3}\,\text{eV}$ symmetron due to the electron magnetic moment. The regions of parameter space excluded at the 95\% confidence level are shown as two-dimensional slices for different values of (a)~$M_\gamma$, (b)~$M_s$, and (c)~$\lambda$. We show constraints only for weak couplings $\lambda \lesssim 1/6$ which are amenable to our perturbative approach. Other approximations are also responsible for moulding the final shape of the shaded regions shown here. These are discussed towards the end of Sec.~\ref{sec:sym_c_matter}, and are primarily responsible for the awkward shapes of the bottom edges.}
\label{fig:sym_c_2}
\end{figure*}
\begin{figure*}
\includegraphics[width=\textwidth]{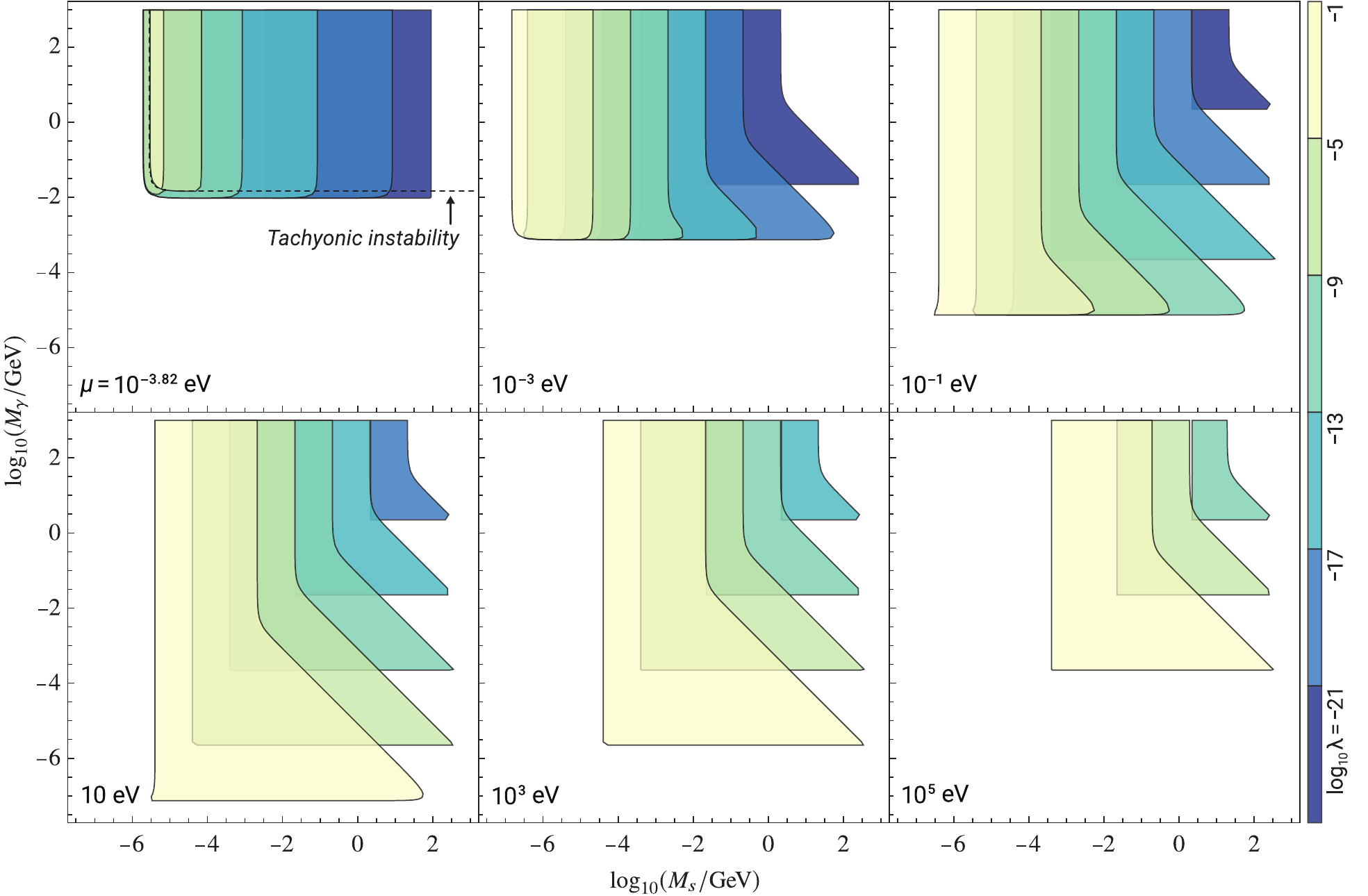}
\caption{Constraints on the symmetron due to the electron magnetic moment for different values of $\mu$. Shaded regions denote the values of the parameters in the $M_s$--$M_\gamma$ plane that are excluded at the 95\% confidence level for each value of $\lambda$. We show constraints only for weak couplings $\lambda \lesssim 1/6$ which are amenable to our perturbative approach. In the $\mu = 10^{-3.82}\,\text{eV}$ panel, the slice for the largest value of $\lambda$ does not extend as far to the left and bottom as the others. This is because the quantum correction terms responsible for this slice suffer from a tachyonic instability near the edges, when $\varphi_0 < 1/\sqrt{3}$ (see the last paragraph of Sec.~\ref{sec:sym_c_matter} for details).}
\label{fig:sym_c_3}
\end{figure*}

One last subtlety must be brought to light. Our calculations for the quantum corrections also fail to hold when the symmetron becomes tachyonic at the center ($m_0^2 < 0$). Rather than signaling any kind of severe pathology with the theory, this merely indicates that we can no longer neglect the spatial variation of $\langle\phi\rangle$ when computing the quantum corrections. As such a calculation is beyond the scope of this paper, we have simply forgone placing constraints when this occurs. Luckily this does not affect the end results much, and explains why the shaded region due to the quantum corrections in Fig.~\ref{fig:sym_c_1}(a) does not extend as far to the left as the cavity shift.

\subsubsection{Photon coupling}

We now discuss the constraints on the symmetron when the photon coupling $M_\gamma$ is included. For an illustrative value of $\mu = 10^{-3}\,\text{eV}$, the region in the $(\lambda, M_s, M_\gamma)$ subspace that is excluded is shown in Fig.~\ref{fig:sym_c_2}. As before, the bottom edges of each shaded region in Figs.~\ref{fig:sym_c_2}(a) and \ref{fig:sym_c_2}(b) correspond to the boundary beneath which the weak coupling limit and, further down, the EFT itself stop being valid. These conditions correspond to Eqs.~\eqref{eq:sym_validity_wc_2} and \eqref{eq:sym_validity_eft}, respectively, and are universal to all experiments.\footnote{More precisely, Eq.~\eqref{eq:sym_validity_wc_2} applies only to experiments probing the quantum nature of the symmetron for which a perturbative calculation is unavoidable, whereas Eq.~\eqref{eq:sym_validity_eft} applies in all cases.} For this reason, the most essential information to be gained from this experiment is encapsulated in the top edges of the shaded regions, which give the lower bound on $\lambda$ that remains viable. Evidently, this lower bound on $\lambda$ increases for fixed $(\mu, M_s)$ as we decrease $M_\gamma$. This information is most efficiently conveyed in a ``top view'' plot as shown in Fig.~\ref{fig:sym_c_2}(c).

To observe the dependence on the symmetron mass, top-view plots for different values of $\mu$ are shown in Fig.~\ref{fig:sym_c_3}. For $\mu \lesssim 10^{-3}\,\text{eV}$, the photon coupling has no noticeable effect, whereas the shapes of the shaded regions are qualitatively similar for all $\mu > 10^{-3}\,\text{eV}$. As we increase $\mu$, the left and bottom edges of each slice in the $M_s$--$M_\gamma$ plane move further left and bottom, owing to the fact that spontaneous symmetry breaking in the cavity can occur for smaller values of $M_s$ and $M_\gamma$. This continues on until about $\mu \sim 10\,\text{eV}$, when the opposite begins to occur and the edges retreat towards the top-right corner of the plot. This happens simply because Eqs.~\eqref{eq:sym_validity_wc_2} and \eqref{eq:sym_validity_eft} break down in larger and larger regions of parameter space as $\mu$ increases. When we reach $\mu \sim 10^8\,\text{eV}$, the theory becomes completely unworkable in the range of $M_s$ and $M_\gamma$ accessible to this experiment, such that no constraint can be placed.

Viewed from this perspective, the shaded regions in the $M_s$--$M_\gamma$ plane for a given value of $\lambda$ strongly resemble the chameleon constraints in the $M_c$--$M_\gamma$ plane of Fig.~\ref{fig:chm_c_1}(a). The effect of the photon coupling only becomes noticeable below a certain value of $M_\gamma$, and the lower bound depends on the specific value of $M_s$. This behavior can of course be traced back to the quantum correction terms, where the photon coupling always appears in tandem with the matter coupling when restricted to leading one-loop order. Recall in the case of the chameleon that we spent no effort illustrating the weak constraints on the photon coupling any further, since they were found to be uncompetitive with those already placed by collider experiments \cite{Brax:2009aw}. The same might be true for the symmetron, although no work has yet been done to translate the bounds and demonstrate this definitively. Indeed, to our knowledge, this paper represents the first attempt at constraining the symmetron's coupling to photons.

\section{Conclusion}
\label{sec:conclusions}

Decades of exceptional work by theorists and experimentalists alike have now verified the accuracy of the Standard Model, and QED in particular, to about one part per trillion. Beyond achieving their original objective, we have shown that precision tests of QED can also be used to place meaningful constraints on the existence of chameleonlike particles (CLPs) that mediate screened fifth forces. In this work, we considered the implications of the precision measurement of the electron's magnetic moment, focusing on two main scalar-induced effects that could arise.

First, the virtual exchange of CLPs generates additional loop corrections to the QED vertex function, since the scalars are assumed to couple to electrons and photons with gravitational strength or greater (see Sec.~\ref{sec:quantum}). This leads to an increase in the intrinsic value of the magnetic moment, which must be constrained to be less than $\sim 10^{-12}$ lest it ruin the remarkable agreement between experiment and the Standard Model prediction. Second, nonlinear self-interactions drive the scalar to form a bubblelike profile within the cylindrical vacuum cavity of the experiment. This scalar profile exerts an additional fifth force on the electron confined to the Penning trap, thus perturbing its energy eigenvalues. A systematic shift of this form can also be used to place constraints, since the magnetic moment is determined experimentally by measuring the transition frequencies between energy levels (see Sec.~\ref{sec:cav}).

Accurate estimates of these effects require knowledge of the value of the scalar field at the center of the cavity, which can only be determined by fully solving the nonlinear field equation. The absence of any known closed-form solution---either approximate or exact---for the case of the cylindrical geometry considered here has led to a somewhat novel, semiempirical approach. It has already been shown that a chameleon in a vacuum cavity satisfies a resonance condition such that its local Compton wavelength is dynamically adjusted to match the size of the cavity \cite{Brax:2007hi}. In this paper, we have shown this explicitly for the case of a plane-parallel cavity by obtaining an approximate, one-dimensional solution. Through well-motivated arguments, the solution to this toy model was then deformed to describe more arbitrary convex cavity shapes. The resulting empirical formula for the central field value is a function of only two free parameters, which are tuned to best fit the full numerical solutions carried out for a small number of points in parameter space (see Sec.~\ref{sec:chm}).

We found that the quantum corrections were able to place a universal bound of $\log_{10}(M_c/\Mp) \gtrsim -16.7$ for the chameleon model, independent of the values of $(\Lambda, n)$. However, for values near $\Lambda \approx 300\,\text{eV}$, the cavity shift dominates to give a much better lower bound of $\log_{10}(M_c/\Mp) \gtrsim -10$. While this part of parameter space is already constrained by other laboratory experiments, the bound determined here represents the tightest constraint yet achieved by an experiment not originally intended to search for fifth forces.

Our results are able to break even more ground for the symmetron (see Sec.~\ref{sec:sym}). Again, a deformation of the one-dimensional solution to a plane-parallel cavity results in an empirical formula with only two free parameters that can be tuned to fit the numerical results with a high degree of accuracy. With this in hand, we saw that the cavity shift places constraints only for a small range of the symmetron mass, $\mu \in [10^{-3.88}, 10^{-3.39}]\,\text{eV}$. This limitation is unsurprising, and is generic to any laboratory experiment that probes the effect of the symmetron's fifth force. When $\mu$ is too small, the associated Compton wavelength is too large such that the symmetron is unable to resolve the size of the vacuum cavity, and thus remains in the symmetry-unbroken phase. On the other end of the spectrum, the fifth force is strongly Yukawa-suppressed when $\mu$ is too large, resulting in a field profile that is essentially flat in the cavity except near the walls.

Nonetheless, the electron magnetic moment has an added advantage over other experiments that have hitherto provided constraints on the symmetron. The quantum corrections are able to yield constraints regardless of the value of $\mu$, provided only that the mass is large enough to enable spontaneous symmetry breaking, and small enough that the effective field theory remains valid. As a result, this experiment has probed, and decisively ruled out, a large and previously unexplored region of parameter space in the range $\mu \in [10^{-3.88}, 10^8]\,\text{eV}$ for couplings ($M_s, M_\gamma$) around the GeV scale.

To conclude, this work provides a clearer picture of the space of CLP models that remain viable in this Universe, now more than ever. Our results also suggest a new direction for future work: While dedicated fifth-force experiments such as atom interferometry and torsion balances may well provide the best sensitivities in a given mass range near the meV scale, it will be interesting to explore other experiments that exploit the quantum nature of the symmetron in the hopes of covering large regions of parameter space more efficiently.

\begin{acknowledgments}
It is a pleasure to thank Clare Burrage for helpful discussions. This work has been partially supported by STFC Consolidated Grants No.~ST/P000673/1 and No.~ST/P000681/1. This work is supported in part by the EU Horizon 2020 research and innovation program under the Marie-Sklodowska Grant No.~690575. This article is based upon work related to the COST Action CA15117 (CANTATA) supported by COST (European Cooperation in Science and Technology). B.E. is supported by a Leverhulme Trust Research Leadership Award. L.K.W. is supported by the Cambridge Commonwealth, European and International Trust, and Trinity College, Cambridge. We also thank the organizers of the \emph{PSI$^2$ DarkMod} and \emph{Dark Energy in the Laboratory} workshops for the conducive environments they have provided during which some of this work was completed.
\end{acknowledgments}

\appendix
\section{One-loop Feynman diagrams}
\label{app:feyn}

In this Appendix, we briefly outline the calculations that lead to the results in Eqs.~\eqref{eq:da_bMbM} and \eqref{eq:da_bMbG}. The steps taken here are all standard techniques; easily found in any introductory Quantum Field Theory textbook. Our conventions follow those of Ref.~\cite{Srednicki}, except that we take the electron to have charge $-e$, meaning our constant $e > 0$.

The electron magnetic moment is determined by computing the renormalized QED vertex function $\Gamma^\mu(p,p')$, where we take $p$ to be the momentum of the incoming fermion, $p'$ to be that of the outgoing fermion, and let $q = p'-p$ be the momentum of the ingoing photon. We have normalized by a factor of $-ie$, such that the tree-level contribution is
\begin{equation*}
\Gamma^\mu_\text{tree-level} = \gamma^\mu.
\end{equation*}
Our theory respects Lorentz invariance, $U(1)$ gauge invariance, and $CP$-symmetry, hence the most general form of this vertex is
\begin{equation}
\label{eq:feyn_vertex}
\Gamma^\mu = F_1(q^2) \gamma^\mu + F_2(q^2) \left( \frac{-i S^{\mu\nu}q_\nu}{m_e}\right)
\end{equation}
when the external fermions are on-shell. In Sec.~\ref{sec:quantum}, recall we defined $S^{\mu\nu} = \frac{i}{4}[\gamma^\mu,\gamma^\nu]$, with the gamma matrices satisfying $\{ \gamma^\mu,\gamma^\nu \} = -2 \eta^{\mu\nu}$. The functions $F_{1,2}$ are typically called the electric and magnetic form factors. The constant electric part $F_1(0)$ is a renormalization of the electron charge, which can be set to $F_1(0) = 1$ exactly in an on-shell renormalization scheme. The constant magnetic part is exactly the anomalous magnetic moment, $F_2(0) = a$.

A scalar field, like the chameleon or symmetron, contributes via three Feynman diagrams to this vertex at the one-loop level, shown in Fig.~\ref{fig:FeynmanDiagrams}. The first of these, in Fig.~\ref{fig:FeynmanDiagrams}(a), involves only the Yukawa-like matter coupling, and we shall refer to this as the Yukawa-type diagram. The remaining diagrams involve the photon coupling, and are sometimes called Barr-Zee-type diagrams \cite{PhysRevLett.65.21,*PhysRevLett.65.2920,Marciano:2016yhf}.

For brevity, we shall soon write integrals over $d$-dimensional loop momenta and over Feynman parameters, respectively, as
\begin{align*}
\int_l &= \int\frac{\text{d}^dl}{(2\pi)^d},
\\
\int_{[n]} &= (n-1)!\int_0^1 \text{d}x_1 \dots \int_0^1\text{d}x_n \delta\left(\sum_i x_i - 1\right).
\end{align*}

\subsection{Yukawa-type diagram}
Standard Feynman rules dictate that the contribution of Fig.~\ref{fig:FeynmanDiagrams}(a) to the vertex function is
\begin{equation*}
i\Gamma^\mu_\text{(a)} = \frac{\beta_m^2m_e^2}{\Mp^2} \int_l
\frac{(-\slashed l - \slashed p' + m_e)\gamma^\mu(-\slashed l - \slashed p + m_e)}{[(l+p')^2 + m_e^2][(l+p)^2 + m_e^2][l^2 + m_0^2]}.
\end{equation*}
Using Feynman parametrization and defining a new integration variable $k = l + x_1 p + x_2 p'$, this becomes
\begin{equation}
i \Gamma^\mu_\text{(a)} = \frac{\beta_m^2m_e^2}{\Mp^2} \int_{[3]}\int_k \frac{N^\mu_\text{(a)}}{(k^2 + D_\text{(a)})^3},
\end{equation}
where the numerator and denominator are
\begin{align}
N^\mu_\text{(a)} &= (-\slashed l - \slashed p' + m_e)\gamma^\mu(-\slashed l - \slashed p + m_e)|_{l = k-x_1p-x_2p'},
\nonumber\\
D_\text{(a)} &= x_1 (1-x_1) p^2 + x_2(1-x_2) p'^2 - 2x_1 x_2 p \cdot p'
\nonumber\\
&\quad+ (x_1 + x_2) m_e^2 + x_3 m_0^2.
\end{align}

As $N^\mu_\text{(a)}$ sits under an integral over all $k$, terms linear in $k$ vanish upon integration. For this same reason, terms quadratic in $k$ will simplify to
\begin{equation}
N^\mu_\text{(a)} \supset \slashed k \gamma^\mu \slashed k = \frac{d-2}{2}k^2 \gamma^\mu,
\end{equation}
where the equality holds only under the integral. The integrand is now a function only of $k^2$, and the loop integral can be performed using the standard formula
\begin{equation}
\int_k \frac{(k^2)^a}{(k^2 + D)^b} = i \frac{\Gamma(b-a-d/2)\Gamma(a+d/2)}{(4\pi)^{d/2}\Gamma(b)\Gamma(d/2)} D^{-(b-a-d/2)}.
\end{equation}
The factor of $i$ on the rhs appears from Wick rotation. The terms in the numerator quadratic in $k$ integrate to give a log-divergent piece proportional to $\gamma^\mu$. This is a contribution only to $F_1$, and is merely a renormalization of the electron charge. For our purposes, the terms of interest sit in the $k$-independent part of $N^\mu_\text{(a)}$. Let us refer to this as
\begin{equation}
N^\mu_\text{(a)} \supset (\slashed Q_1 + m_e)\gamma^\mu(\slashed Q_2 + m_e) \eqcolon n^\mu_\text{(a)},
\end{equation}
where $Q_1 = x_1p - (1-x_2)p'$ and $Q_2 = x_2p' - (1-x_1)p$.

We can simplify this further, since we only care about $\Gamma^\mu$ on-shell. This is when it sits in an S-matrix element of the form $\overline u(p') \Gamma^\mu u(p) A_\mu(q)$. The external photon $A_\mu$ is classical and off-shell, corresponding to the large magnetic field in the cavity. Writing $u \equiv u(p)$ and $\ovl u' \equiv \ovl u(p')$, the momentum eigenstates of the fermion satisfy
\begin{equation}
\label{eq:feyn_up}
\slashed p u = -m_e u, \quad
\ovl u' \slashed p' = - m_e\ovl u'.
\end{equation}
The spin indices on $u,\ovl u'$ have been suppressed as they are not essential here.

The name of the game now is to reorder the terms in $n^\mu_\text{(a)}$ by using the anticommutation relations such that use can be made of Eq.~\eqref{eq:feyn_up}. The end result is
\begin{align}
n^\mu_\text{(a)} &= [4m_e^2 - (1-x_3)^2m_e^2 - x_1 x_2 q^2]\gamma^\mu
\nonumber\\
&\quad- m_e (1-x_3)(1+x_3)(p'+p)^\mu 
\nonumber\\
&\quad+ m_e [(2x_1 - x_1^2)-(2x_2-x_2^2)](p'-p)^\mu.
\end{align}
The final line changes sign under the exchange $x_1 \leftrightarrow x_2$, whereas the remainder of the integrand is unchanged. Hence, this term vanishes upon integration over the Feynman parameters. We then use the Gordon identity
\begin{equation}
\label{eq:feyn_Gordon}
\ovl u' (p' + p)^\mu u = \ovl u'(2 m_e \gamma^\mu + 2 i S^{\mu\nu} q_\nu)u
\end{equation}
to recast the second line into a form comparable to Eq.~\eqref{eq:feyn_vertex}. As before, terms proportional to $\gamma^\mu$ contribute only to $F_1$, and are not interesting to us. After integration over $k$ with $d=4$, the contribution to $F_2$ is
\begin{equation}
F_\text{2,(a)}(q^2) = \beta_m^2 \left(\frac{m_e}{4\pi\Mp}\right)^2 \int_{[3]}\frac{(1-x_3)(1+x_3)}{D_\text{(a)}(q^2)/m_e^2}.
\end{equation}
Setting $p^2 = p'^2 = -m_e^2$ and $q^2 = 0$, the denominator $D_\text{(a)}$ when on-shell takes the form
\begin{equation*}
D_\text{(a)}(0) = (1-x_3)^2 m_e^2 + x_3 m_0^2.
\end{equation*}
Integrating over $x_1$ and $x_2$, and renaming $x_3$ as just $x$ returns the desired result in Eq.~\eqref{eq:da_bMbM}.

\subsection{Barr-Zee-type diagrams}
We can now repeat the same steps for the remaining diagrams. The contribution from Fig.~\ref{fig:FeynmanDiagrams}(b) is
\begin{equation*}
i\Gamma^\mu_\text{(b)} = \frac{\beta_m\beta_\gamma m_e}{\Mp^2} \int_l
\frac{(\slashed l + m_e) \gamma_\nu [(l+p)^\mu q^\nu - q\cdot(l+p) \eta^{\mu\nu}]}{[l^2 + m_e^2][(l+p')^2 + m_0^2](l+p)^2}.
\end{equation*}
Defining $k$ exactly as before, this can be rewritten as
\begin{equation}
i\Gamma^\mu_\text{(b)} = \frac{\beta_m\beta_\gamma m_e}{\Mp^2} \int_{[3]}\int_k
\frac{N^\mu_\text{(b)}}{(k^2 + D_\text{(b)})^3}.
\end{equation}
In terms of a constant matrix $\Delta_{\alpha\beta}^{\mu\nu} = \delta_\alpha^\mu \delta_\beta^\nu - \eta_{\alpha\beta}\eta^{\mu\nu}$, the numerator and denominator are
\begin{align}
N^\mu_\text{(b)} &= (\slashed l + m_e) \gamma_\nu (l+p)^\alpha q^\beta \Delta_{\alpha\beta}^{\mu\nu}|_{l = k-x_1 p - x_2 p'},
\nonumber\\
D_\text{(b)} &= x_1(1-x_1) p^2 + x_2(1-x_2)p'^2 - 2x_1x_2 p\cdot p'
\nonumber\\
&\quad+ x_2 m_0^2 + x_3 m_e^2.
\end{align}

The terms in $N^\mu_\text{(b)}$ linear in $k$ vanish upon integration, so we need again only pay attention to the terms quadratic in and independent of $k$. The former simplifies to
\begin{equation}
N^\mu_\text{(b)} \supset \slashed k \gamma_\nu k^\alpha q^\beta \Delta_{\alpha\beta}^{\mu\nu} = -i\frac{4k^2}{d} S^{\mu\nu} q_\nu.
\end{equation}
Again, we note that the second equality holds only under the integral. This contributes a log-divergent piece to $F_2$, which we regulate by performing the integral in $d=4-\epsilon$ dimensions. In the $\overline{\text{MS}}$~scheme, we keep the coupling strengths $\beta_i$ dimensionless by pulling out an explicit mass dependence,
\begin{equation*}
\beta_i \to \beta_i \tilde\mu^{\epsilon/2},
\end{equation*}
where the arbitrary mass scale $\mu$ is defined via $\mu^2 = 4\pi e^{-\gamma_E} \tilde\mu^2$ in terms of the Euler-Mascheroni constant $\gamma_E$. Performing the momentum integral, we get
\begin{equation}
F_\text{2,(b)}(q^2) \supset \beta_m\beta_\gamma \left(\frac{m_e}{4\pi\Mp}\right)^2 \int_{[3]}
\left[ \frac{2}{\epsilon} + \log\left(\frac{\mu^2}{D_\text{(b)}}\right) \right].
\end{equation}
The $\mathcal O(1/\epsilon)$ term is removed by an appropriate counterterm (discussed in Sec.~\ref{sec:quantum_ren}). Working on-shell, the denominator is
\begin{equation*}
D_\text{(b)}(0) = x_3^2 m_e^2 + x_2 m_0^2.
\end{equation*}
We integrate over $x_1$, rename $x_3 = x$, and make the change of variables $x_2 = (1-x)y$ to bring this into the form
\begin{equation}
F_\text{2,(b)}(0) \supset 2\beta_m\beta_\gamma\left(\frac{m_e}{4\pi\Mp}\right)^2
\left[ \log\left(\frac{\mu}{m_e}\right) + I_2\left(\frac{m_0}{m_e}\right)\right],
\end{equation}
where the integral $I_2$ is given in Eq.~\eqref{eq:feyn_integral_2}. This is half of the desired result in Eq.~\eqref{eq:da_bMbG}. Unsurprisingly, the other half comes from evaluating Fig.~\ref{fig:FeynmanDiagrams}(c), which turns out to give exactly the same contribution as Fig.~\ref{fig:FeynmanDiagrams}(b) when on-shell.

Given we have already obtained the desired outcome, we are left to show that the terms independent of $k$ in $N^\mu_\text{(b)}$ and $N^\mu_\text{(c)}$ do not contribute to $F_2(0)$. As the manipulations are near identical, we shall describe the general procedure only for $N^\mu_\text{(b)}$. Its $k$-independent terms are
\begin{equation}
N^\mu_\text{(b)} \supset (\slashed l + m_e) \gamma_\nu (l+p)^\alpha q^\beta \Delta_{\alpha\beta}^{\mu\nu}|_{l = -x_1 p - x_2 p'} \eqcolon n^\mu_\text{(b)}.
\end{equation}

Once again, we begin by employing the anticommutation relations to reorder terms until Eq.~\eqref{eq:feyn_up} can be judiciously applied. We find
\begin{equation}
n^\mu_\text{(b)} = \left(2 x_2 - \frac{x_3^2}{2}\right) q^2 m_e \gamma^\mu + 2 x_1 x_2 q^2 i S^{\mu\nu}q_\nu,
\end{equation}
having also used the Gordon identity in Eq.~\eqref{eq:feyn_Gordon}. Its contributions to both $F_1$ and $F_2$ are proportional to $q^2$, so do not affect either the electric charge or the anomalous magnetic moment.

\section{Numerically\\ accessible region}
\label{app:approx}

Several approximations had to be made in this paper in order that the numerical work remain tractable. These approximations are valid only in certain regions of parameter space, outside of which our calculations are not trustworthy. In this Appendix, we briefly discuss how we estimate the boundaries of this so-called numerically accessible region (NAR).

\subsection{Chameleon}

For the chameleon model, we assumed the zero-skin-depth (ZSD) and perfect-vacuum (PV) approximations, which correspond to Eqs.~\eqref{eq:chm_approx_zsd} and \eqref{eq:chm_approx_pv}, respectively. Here, we make these statements more precise.

The ZSD approximation is implemented numerically by fixing the field at the value that minimizes its local effective potential once it reaches the walls. In reality, the field cannot achieve this instanteously, but rather decays to the minimum within a distance set by its Compton wavelength $m_\infty^{-1}$. This fact is compatible with the ZSD approximation provided $m_\infty^{-1}$ cannot be resolved by our numerical code. Following Ref.~\cite{Elder:2016yxm}, we therefore require that the Compton wavelength be at least an order of magnitude smaller than the numerical grid spacing for this approximation to be valid,
\begin{equation}
\label{eq:nar_chm_zsd}
m_\infty^{-1} < \frac{l_\text{grid}}{10}.
\end{equation}
We used a grid spacing $l_\text{grid} = 0.1\,\text{mm}$ in all three spatial directions for the chameleon. Naturally, the Compton wavelength in the vacuum chamber $m_0^{-1}$ is much larger than $l_\text{grid}$; hence, satisfying Eq.~\eqref{eq:nar_chm_zsd} is sufficient to guarantee we also satisfy Eq.~\eqref{eq:chm_approx_zsd}.

A different tactic is required to determine when the PV approximation breaks down. To do so, we recast Eq.~\eqref{eq:chm_approx_pv} as an inequality
\begin{equation}
\label{eq:nar_chm_pv}
\frac{\rho_\text{cav}}{M_c} + \frac{\rho_\text{em}}{M_\gamma} < \epsilon_\textsc{pv} \frac{n \Lambda^{4+n}}{\phi_0^{n+1}},
\end{equation}
and shall utilize the approximate one-dimensional solutions obtained in Sec.~\ref{sec:chm_1D} to determine an appropriate value for~$\epsilon_\textsc{pv}$. This is done as follows: Unlike in Sec.~\ref{sec:chm_1D}, we now solve Eq.~\eqref{eq:chm_solve_phi0} for $\phi_0$ without making the PV approximation. This is only possible numerically. We substitute in
\begin{equation*}
V'_0 = - \frac{n\Lambda^{4+n}}{\phi_0^{n+1}}(1 - \epsilon_\textsc{pv})
\end{equation*}
into the equation and vary the value $\epsilon_\textsc{pv}$, observing how the solution $\phi_0$ changes. For the chameleon, we are typically interested in constraining the order of magnitude of the coupling scales $M_i \in \{ M_c, M_\gamma \}$ for given values of $(\Lambda,n)$. Thus, our criterion is to tolerate a value for $\epsilon_\textsc{pv}$ that leads to a change of at most $\pm 0.1$ in the value of the constraint on $\log_{10} (M_i/\Mp)$. We find that choosing
\begin{equation*}
\epsilon_\textsc{pv} \simeq 0.34
\end{equation*}
ensures we satisfy this criterion. As Eq.~\eqref{eq:nar_chm_pv} makes no specific reference to the geometry of the problem, and as the value for~$\epsilon_\textsc{pv}$ is small, we expect this result to be a good estimate also for the two-dimensional cylindrical case.
 
The boundaries defined by Eqs.~\eqref{eq:nar_chm_zsd} and \eqref{eq:nar_chm_pv} demarcate the region of parameter space outside of which calculations for the cavity shift can no longer be trusted. However, the quantum corrections calculated in Sec.~\ref{sec:quantum} depend much more weakly on $\phi_0$. As can be seen in Eqs.~\eqref{eq:da_bMbM} and \eqref{eq:da_bMbG}, $\phi_0$ enters only through the ratio $m_0/m_e$, which remains small long after both the ZSD and PV approximations break down. Consequently, the constraints from the quantum corrections hold well beyond the NAR. We can determine when the approximation $m_0/m_e = 0$ finally breaks down in a similar fashion. Again tolerating a change of at most $\pm 0.1$ in $\log_{10} (M_i/\Mp)$, we require that the integrals $I_{1,2}(\eta)$, given in Eq.~\eqref{eq:feyn_integrals_closedform}, decrease by at most 40\%. For small $\eta = m_0/m_e$, it suffices to impose this just on $I_1(\eta)$. This translates to the condition
\begin{equation}
\frac{m_0}{m_e} < 0.31.
\end{equation}
Of course, we do not have a good way to determine the value of $m_0$ now that we are outside the NAR. A conservative estimate is to replace $m_0$ above with its maximum possible value, which is when $\phi_0$ minimizes the effective potential in the cavity. For $n=1$ and $\Lambda = 2.4\,\text{meV}$, this puts the boundary at
\begin{equation*}
\log_{10}(M_c/\Mp) \gtrsim -31.4, \quad
\log_{10}(M_\gamma/\Mp) \gtrsim -24.0.
\vspace{0.7em}
\end{equation*}
Increasing either $\Lambda$ or $n$ pushes the boundary even further out. Couplings much larger than this boundary are already in tension with classical fifth force tests and constraints from particle colliders \cite{Burrage:2017qrf,Burrage:2016bwy}, hence we can be assured that the constraints from quantum corrections hold throughout the region plotted in Fig.~\ref{fig:chm_c_1}.

\subsection{Symmetron}

Only the ZSD approximation is needed for the symmetron. The condition, originally given in Eq.~\eqref{eq:sym_zsd}, can be replaced by the inequality
\begin{equation}
\label{eq:nar_sym_zsd}
\mu_\infty^{-1} < \frac{l_\text{grid}}{10}.
\end{equation}
Like the chameleon, we again require that the Compton wavelength of the symmetron be smaller than the numerical grid spacing. Unlike the chameleon, however, a finer grid with $l_\text{grid} = 0.05\,\text{mm}$ in all three spatial directions was required to ensure convergence, especially for solutions with values of $\varphi_0$ close to zero.

Note that Eq.~\eqref{eq:nar_sym_zsd} need only be imposed for symmetron masses in the range $\mu = [10^{-3.88}, 10^{-3.39}]\,\text{eV}$. For smaller values, the symmetron remains in the symmetry-unbroken phase where no constraint can be placed. For larger values of $\mu$, the symmetron quickly reaches the local maximum $\varphi_0 = 1$ inside the cavity irrespective of what is happening in and beyond the walls. 
\bibliography{g}
\end{document}